\documentclass{aa}

\usepackage[T1]{fontenc}
\usepackage[latin1]{inputenc}
\usepackage{graphicx}
\usepackage{wasysym}
\usepackage{natbib}
\usepackage{txfonts}

\newcommand{\eq}[1]{\begin{equation}  #1 \end{equation}}

\newcommand{\eqa}[1]{\begin{eqnarray}   #1 \end{eqnarray}}

\newcommand{\br}[1]{\left( #1 \right)}
\newcommand{\bc}[1]{\left\{ #1 \right\}}
\newcommand{\bb}[1]{\left[ #1 \right]}

\newcommand{\nn}{\nonumber}

\newcommand{\dd}{{\rm d}}
\newcommand{\expo}[1]{~{\rm e}^{ #1 }}
\newcommand{\vek}[1]{\mbox{\boldmath $#1$}}

\begin{document}

\title{Intrinsic alignment boosting}
\subtitle{Direct measurement of intrinsic alignments in cosmic shear data}

\author{B. Joachimi \and P. Schneider}

\offprints{B. Joachimi,\\
    \email{joachimi@astro.uni-bonn.de}
}

\institute{Argelander-Institut f\"ur Astronomie (AIfA), Universit\"at Bonn, Auf dem H\"ugel 71, 53121 Bonn, Germany}

\date{Received 22 March 2010 / Accepted 18 April 2010}

\abstract{}
{Intrinsic alignments constitute the major astrophysical systematic for cosmological weak lensing surveys. We present a purely geometrical method with which one can study gravitational shear-intrinsic ellipticity correlations directly in weak lensing data.}
{Linear combinations of second-order cosmic shear measures are constructed such that the intrinsic alignment signal is boosted while suppressing the contribution by gravitational lensing. We then assess the performance of a specific parametrisation of the weights entering these linear combinations for three representative survey models. Moreover a relation between this boosting technique and the intrinsic alignment removal via nulling is derived.}
{For future all-sky weak lensing surveys with photometric redshift information the boosting technique yields statistical errors on model parameters of intrinsic alignments whose order of magnitude is compatible with current constraints determined from indirect measurements. Parameter biases due to a residual cosmic shear signal are negligible in case of quasi-spectroscopic redshifts and remain sub-dominant for typical values of the photometric redshift scatter. We find good agreement between the performance of the intrinsic alignment removal based on the boosting technique and standard nulling methods, both reducing the cumulative signal-to-noise by about a factor of 6, which possibly indicates a fundamental limit in the separation of lensing and intrinsic alignment signals.}
{}


\maketitle

\section{Introduction}
\label{sec:intro}

Weak gravitational lensing of the large-scale structure is going to be one of the major cosmological probes contributing to reveal the properties of dark matter and dark energy in the near future \citep{albrecht06,peacock06}. Within the past decade the method has evolved from its first detections \citep{bacon00,kaiser00,vwaer00,wittman00} to maturity, nowadays yielding statistical constraints which are compatible to other probes (for recent measurements see e.g. \citealp{benjamin07}; \citealp{fu07}; \citealp{schrabback09}; for a recent review see \citealp{munshi08}). Planned surveys measuring weak lensing on cosmological scales, or cosmic shear in short, include Pan-STARRS\footnote{\texttt{http://pan-starrs.ifa.hawaii.edu}}, KIDS\footnote{\texttt{http://www.astro-wise.org/projects/KIDS}}, DES\footnote{\texttt{https://www.darkenergysurvey.org}}, LSST\footnote{\texttt{http://www.lsst.org}}, and Euclid\footnote{\texttt{http://sci.esa.int/science-e/www/area/\\index.cfm?fareaid=102}}. 

The increasingly large statistical power of these surveys demands a more and more thorough treatment of systematic errors. The major astrophysical contamination to cosmic shear is constituted by the intrinsic alignment of galaxies. To infer cosmic shear information from the correlation of galaxy ellipticities, it is usually assumed that the intrinsic shapes of galaxy images are purely random, so that only the desired correlations of gravitational shear (GG in the following) remain. However, due to interactions with the surrounding matter structure, galaxy shapes can intrinsically align, causing correlations between the intrinsic ellipticities of galaxies (II hereafter). Moreover matter can influence the shape of a close-by galaxy via tidal forces and at the same time contribute to the lensing signal of a background galaxy, thereby producing gravitational shear-intrinsic ellipticity correlations (GI hereafter).

Intrinsic alignments have been subject to extensive studies, both analytical and using simulations \citep{croft00,heavens00,lee00,pen00,catelan01,crittenden01,jing02,mackey02,bosch02,hirata04,heymans06,bridle07a,semboloni08,okumura08,okumura09,brainerd09}. Results vary widely, but are mostly consistent with a contamination of the order $10\,\%$ by both II and GI signals for future weak lensing surveys, which can lead to serious biases on cosmological parameters if left untreated \citep[e.g.][]{bridle07}. Intrinsic alignments depend intricately on the formation and evolution of galaxies within their dark matter environment, so that models cannot be expected to develop far beyond the current crude level in the near future. For the most recent advancement in intrinsic alignment modelling see \citet{schneiderm09}.

Using uncertain models of limited accuracy for assessing systematics in statistical analyses is risky \citep{kitching08}. Therefore observational data which can put limits on the possible range of intrinsic alignment signals are highly warranted. It should be noted that in principle intrinsic alignments constitute an interesting cosmological signal worth investigating, shedding light onto the interaction between galaxies, their haloes, and the large-scale structure. Both II and GI correlations have been subject to investigations in several data sets \citep{brown02,heymans04,mandelbaum06,hirata07,brainerd09,mandelbaum09}, results ranging from null to significant detections, depending strongly on the type and colour of galaxies considered. 

However, none of these observations were direct measurements of intrinsic alignments for the galaxy populations and redshifts which are most interesting for cosmic shear because in those cases the shear signal clearly dominates the correlations of galaxy ellipticities. While the II signal is observed at small redshifts where cosmic shear is negligible, the GI term is usually inferred from cross-correlations between galaxy number densities and ellipticities in samples with spectroscopic redshifts. The latter approach requires the assumption of a simple form of the galaxy bias, which is of limited accuracy and inapplicable on small scales. If one wishes to analyse larger galaxy samples for which only photometric redshift information is available, further signals such as galaxy-galaxy lensing contribute and need to be modelled carefully (see \citealp{bernstein08,joachimi10} for an overview on the types of signals contributing to correlations between galaxy number density and ellipticity).

The II signal is less of a concern because, in order to intrinsically align, a pair of galaxies has to have interacted physically, and hence to be both close on the sky and in redshift. This fact can be used to remove II correlations \citep{king02,king03,heymans03,takada04b}, partly in a fully model-independent way with only marginal loss of statistical power if precise redshift information is available. The GI signal is not restricted to physically close pairs of galaxies, but it can also be eliminated in a purely geometrical way via nulling techniques \citep{joachimi08b,joachimi09}. However, a considerable loss of cosmological information is inherent to nulling, and hence, it is still desirable to have a reliable model of GI correlations at one's disposal to be used with other methods controlling this systematic \citep{king05,bridle07,bernstein08,zhang08,joachimi10}.

In the following we will develop a model-independent technique to extract the GI signal from a cosmic shear data set, thereby allowing for direct measurements of GI correlations on the most relevant galaxy samples. This \lq GI boosting\rq\ approach can be regarded as complementary to nulling both in its purpose and in its implementation. Analogous to the nulling technique, we will construct linear combinations of second-order cosmic shear measures, making only use of the well-known characteristic redshift dependence of the GI and GG terms.

This paper is organised as follows. In Sect.$\,$\ref{sec:method} we present the principle of GI boosting and derive general conditions, which are used in Sect.$\,$\ref{sec:weights} to explicitly construct weight functions for the boosting transformation of the cosmic shear signal. Section \ref{sec:modelling} details the modelling which we apply in Sect.$\,$\ref{sec:performance} to assess the performance of the boosting technique. In Sect.$\,$\ref{sec:nulling} we construct a method to remove GI correlations based on the GI boosting technique and investigate the relation between the new approach and the standard nulling method of \citet{joachimi08b,joachimi09}, before we summarise and conclude in Sect.$\,$\ref{sec:conclusions}.

\section{Method}
\label{sec:method}

\subsection{Basic relations}
\label{sec:basicrelations}

We will base our technique on a tomographic cosmic shear data set, i.e. correlations of galaxy ellipticities which are in addition split into subsamples according to the available redshift information. Analogous to the nulling technique the method outlined in the following does not affect angular scales, so that we can without loss of generality use tomographic power spectra as our two-point cosmic shear measures. For an overview on the basics of cosmic shear see e.g. \citet{schneider06} whose notation we mostly follow.

The convergence power spectrum of cosmic shear, correlating two galaxy samples $i$ and $j$, reads
\eq{
\label{eq:GGdef}
P_{\rm GG}^{(ij)}(\ell) = \frac{9H_0^4 \Omega_{\rm m}^2}{4 c^4} \!\! \int^{\chi_{\rm hor}}_0 \!\!\!\!\!\! \dd \chi\; g^{(i)}(\chi)\, g^{(j)}(\chi)\, \bc{1+z(\chi)}^2 P_{\delta} \br{\frac{\ell}{\chi},\chi}\;,
}
where $P_{\delta}$ is the three-dimensional matter power spectrum, $\ell$ the angular frequency, and $z$ the redshift. The integration runs over all comoving distances $\chi$ up to the comoving distance horizon $\chi_{\rm hor}$. Moreover we have introduced the lensing efficiency 
\eq{
\label{eq:lenseff}
g^{(i)}(\chi) = \int_\chi^{\chi_{\rm hor}} \dd \bar{\chi}\, p^{(i)}(\bar{\chi})\, \br{1 - \frac{\chi}{\bar{\chi}}}\;,
}
where $p^{(i)}(\chi)$ is the probability distribution of comoving distances for galaxy sample $i$. Note that we assume a spatially flat universe throughout. Similar to (\ref{eq:GGdef}), one can define a tomographic power spectrum of shear-ellipticity correlations \citep[for details see e.g.][]{hirata04},
\eqa{
\nn
P_{\rm GI}^{(ij)}(\ell) &=& \frac{3H_0^2 \Omega_{\rm m}}{2 c^2} \int^{\chi_{\rm hor}}_0 \dd \chi\; \br{p^{(i)}(\chi)~g^{(j)}(\chi) + g^{(i)}(\chi)~p^{(j)}(\chi)}\\ 
\label{eq:GIdef}
&& \hspace*{1cm} \times\; \bc{1+z(\chi)} \chi^{-1} P_{\delta {\rm I}} \br{\frac{\ell}{\chi},\chi}\;, 
}
where $P_{\delta {\rm I}}$ denotes the three-dimensional cross-power spectrum between matter density contrast and intrinsic shear field\footnote{The intrinsic shear is defined as the correlated part of the intrinsic ellipticity of a galaxy image \citep[e.g.][]{hirata04,joachimi10}. One can then proceed to construct an intrinsic shear field by assigning to every point in space the intrinsic shear a galaxy would have at this position. For instance, if the intrinsic alignment model of \citet{catelan01} held true, this could simply be done by computing the quadrupole of the local gravitational field.}. Only one of the terms in (\ref{eq:GIdef}) is non-vanishing unless the probability distributions overlap. As II correlations can readily be removed before applying a treatment of the GI signal, we neglect them in this work, so that the total power spectrum, i.e. the actual observable in our study, is given by
\eq{
\label{eq:pobs}
P_{\rm obs}^{(ij)}(\ell) = P_{\rm GG}^{(ij)}(\ell) + P_{\rm GI}^{(ij)}(\ell)\;.
}
A discussion on how II correlations affect the boosting technique is provided in Sect.$\,$\ref{sec:conclusions}.

To derive expressions for the transformed signals, we assume that precise redshift, or equivalently distance, information is available, so that the survey can be sliced into thin tomographic bins. One can then approximate $p^{(i)}(\chi) \approx \delta_{\rm D}(\chi - \chi_i)$, where $\chi_i$ is an appropriately chosen comoving distance in bin $i$. Here $\delta_{\rm D}$ denotes the Dirac delta distribution. The lensing efficiency (\ref{eq:lenseff}) can then be written in the form
\eq{
\label{eq:lensefftransition}
g^{(j)}(\chi_i) \rightarrow g(\chi_j,\chi_i) \equiv \left\{ 
\begin{array}{ll}
1 - \frac{\chi_i}{\chi_j}    &~~\mbox{if}~~ \chi_i < \chi_j\\
0                            &~~\mbox{else}\;.  
\end{array} \right.
}
With these approximations the power spectra (\ref{eq:GGdef}) and (\ref{eq:GIdef}) turn into
\eqa{
\label{eq:GGapprox}
P_{\rm GG}(\chi_i,\chi_j,\ell) &=&  \frac{9H_0^4 \Omega_{\rm m}^2}{4 c^4} \int^{{\rm min}(\chi_i,\chi_j)}_0 \!\!\!\!\!\!\!\! \dd \chi\; g(\chi_i,\chi)\; g(\chi_j,\chi)\\ \nn
&& \hspace*{1cm} \times\; \bc{1+z(\chi)}^2 P_{\delta} \br{\frac{\ell}{\chi},\chi}\;;\\ 
\label{eq:GIapprox}
P_{\rm GI}(\chi_i,\chi_j,\ell) &=& \frac{3H_0^2 \Omega_{\rm m}}{2 c^2}\; \Biggl\{ g(\chi_j,\chi_i)\; \frac{1+z(\chi_i)}{\chi_i}\; P_{\delta {\rm I}} \br{\frac{\ell}{\chi_i},\chi_i}\\ \nn
&& \hspace*{.5cm} +\; g(\chi_i,\chi_j)\; \frac{1+z(\chi_j)}{\chi_j}\; P_{\delta {\rm I}} \br{\frac{\ell}{\chi_j},\chi_j} \Biggr\}\;,
}
where the dependence of the power spectra on the comoving distances of the two galaxy samples involved was made explicit. Note that if $\chi_i < \chi_j$, only the first term contributes to $P_{\rm GI}(\chi_i,\chi_j,\ell)$ whereas for $\chi_i > \chi_j$ only the second term is non-zero.

\subsection{Signal transformation}
\label{sec:trafo}

We seek to find linear combinations of tomographic second-order cosmic shear measures such that in the resulting measures the cosmic shear signal is largely suppressed with respect to the GI signal. The starting point is analogous to the nulling technique as outlined by \citet{joachimi08b}. We define transformed power spectra as
\eq{
\label{eq:defpi}
\Pi_{\rm obs}^{(i)}(\ell) \equiv \int_{\chi_{\rm min}}^{\chi_{\rm hor}} \dd \chi\; B^{(i)}(\chi)\; P_{\rm obs}(\chi_i,\chi,\ell)\;,
}
where $B^{(i)}(\chi)$ is a weight function yet to be determined. Note that (\ref{eq:defpi}) holds also for both the GG and GI contributions individually as the observed power spectrum is a linear superposition of the two, see (\ref{eq:pobs}). We will investigate two different choices for the lower boundary of the integration $\chi_{\rm min}$ in this work. To construct the boosting technique, we choose the maximum range $\chi_{\rm min}=0$ whereas in Sect.$\,$\ref{sec:variant2} we will set $\chi_{\rm min}=\chi_i$ instead.

Inserting (\ref{eq:GIapprox}) into the definition (\ref{eq:defpi}), one finds that
\eqa{
\label{eq:modGI}
\Pi^{(i)}_{\rm GI}(\ell) &=& \frac{3H_0^2 \Omega_{\rm m}}{2 c^2}\; \int_0^{\chi_{\rm hor}} \dd \chi\; B^{(i)}(\chi)\; \Biggl\{ g(\chi,\chi_i)\; \frac{1+z(\chi_i)}{\chi_i}\\ \nn
&& \hspace*{0.5cm} \times\; P_{\delta {\rm I}} \br{\frac{\ell}{\chi_i},\chi_i} + g(\chi_i,\chi)\; \frac{1+z(\chi)}{\chi}\; P_{\delta {\rm I}} \br{\frac{\ell}{\chi},\chi} \Biggr\}\\ \nn
&=& \frac{3H_0^2 \Omega_{\rm m}}{2 c^2}\; G^{(i)}(\chi_i)\; \frac{1+z(\chi_i)}{\chi_i}\; P_{\delta {\rm I}} \br{\frac{\ell}{\chi_i},\chi_i}\\ \nn
&& \hspace*{2.5cm} + \int_0^{\chi_i} \dd \chi\; B^{(i)}(\chi)\; P_{\rm GI}(\chi,\chi_i,\ell)\;,
}
where we defined the function
\eq{
\label{eq:Bcondition}
G^{(i)}(\chi) \equiv \int_\chi^{\chi_{\rm hor}} \dd \bar{\chi}\; B^{(i)}(\bar{\chi}) \br{1 - \frac{\chi}{\bar{\chi}}}\;.
}
Note that the integration absorbed into $G^{(i)}(\chi)$ starts at $\chi$, which corresponds to a lower boundary of $\chi_i$ in the integral over the first term in (\ref{eq:modGI}). This can be done because $g(\chi,\chi_i)$ vanishes for $\chi < \chi_i$, see (\ref{eq:lensefftransition}). Likewise, $\chi < \chi_i$ holds for the second term in (\ref{eq:modGI}) to be non-zero, so that the upper boundary of this integration is changed to $\chi_i$. In addition (\ref{eq:GIapprox}), with only its first term non-vanishing, can be inserted. The first term in the final expression of (\ref{eq:modGI}) is generated by GI correlations originating from matter at the distance $\chi_i$. Note that in the approximation of thin redshift slices which we are working in this term is just a rescaled version of (\ref{eq:GIapprox}). Due to our choice $\chi_{\rm min}=0$, the transformed GI signal receives a further contribution from shear-ellipticity correlations generated at $\chi < \chi_i$, collected into the second term of (\ref{eq:modGI}).

Transforming the lensing signal analogously by plugging (\ref{eq:GGapprox}) into (\ref{eq:defpi}), one arrives at
\eqa{
\label{eq:modGG}
\Pi^{(i)}_{\rm GG}(\ell) &=& \frac{9H_0^4 \Omega_{\rm m}^2}{4 c^4} \int_0^{\chi_i} \dd \chi \int_{\chi}^{\chi_{\rm hor}} \!\!\!\! \dd \bar{\chi}\; B^{(i)}(\bar{\chi}) \br{1 - \frac{\chi}{\bar{\chi}}}\\ \nn
&& \hspace*{2.5cm} \times\; \br{1 - \frac{\chi}{\chi_i}} \bc{1+z(\chi)}^2 P_\delta \br{\frac{\ell}{\chi},\chi}\\ \nn
&=& \frac{9H_0^4 \Omega_{\rm m}^2}{4 c^4} \int_0^{\chi_i} \!\!\! \dd \chi \br{1 - \frac{\chi}{\chi_i}} G^{(i)}(\chi) \bc{1+z(\chi)}^2 P_\delta \br{\frac{\ell}{\chi},\chi}\;.
}
Again, (\ref{eq:Bcondition}) was used to produce the final expression. The conditions $\chi < \chi_i$ and $\bar{\chi} > \chi$, imposed by (\ref{eq:lensefftransition}), result in the upper boundary of the first and the lower boundary of the second integral, respectively. The transformed cosmic shear signal thus depends on the form of $G^{(i)}(\chi)$ in the interval between 0 and $\chi_i$. To suppress the GG signal, $G^{(i)}(\chi)$ should be chosen such that the integral in the final expression of (\ref{eq:modGG}) is close to zero while at the same time $G^{(i)}(\chi_i)$ has to be comparatively large to boost the GI contribution, see (\ref{eq:modGI}).

In reality line-of-sight information will not be available in terms of comoving distances, but rather in terms of the observable redshift. Furthermore the galaxy redshift distributions will have a finite width and also overlap due to scatter, in particular if only photometric redshift information is available as will be the case for the vast majority of galaxies in future cosmic shear surveys. To arrive at a practical prescription for constructing the transformed power spectra, we therefore change the integration variable in (\ref{eq:defpi}) to redshift and subsequently discretise the integral, yielding
\eq{
\label{eq:defpidiscrete}
\Pi_{\rm obs}^{(i)}(\ell) \approx \sum_{j=j_{\rm min}}^{N_z} B^{(i)}(\chi(z_j))\; P_{\rm obs}^{(ij)}(\ell)\; \chi'(z_j)\; \Delta z_j\;,
}
where $\chi'(z)$ is the derivative of comoving distance with respect to redshift, and $\Delta z_j$ is the width of redshift bin $j$. In total $N_z$ galaxy samples are available for study. Here and in the following we identify $z_i \equiv z(\chi_i)$. The condition $\chi_{\rm min}=0$ used for the boosting technique translates into $j_{\rm min}=1$.

\subsection{Solving for the weight function}
\label{sec:solveweight}

In the foregoing section we saw that the GI signal can be boosted, and the GG signal at the same time suppressed, by formulating conditions on the function $G^{(i)}(\chi)$. Via its defining equation (\ref{eq:Bcondition}) it is related to the weight function $B^{(i)}(\chi)$ that enters the transformation (\ref{eq:defpi}). Hence, to obtain a boosting transformation, one has to solve (\ref{eq:Bcondition}) for $B^{(i)}(\chi)$ for a given function $G^{(i)}(\chi)$.

We begin by noting that (\ref{eq:Bcondition}) is a Volterra integral equation of the first kind. It has a kernel that is linear in the integration variable, so that one can readily solve for the weight function by differentiating twice, resulting in
\eq{
\label{eq:volterrasolution}
B^{(i)}(\chi) = \chi\; \frac{\dd^2 G^{(i)}(\chi)}{\dd \chi^2}\;.
}
We have found the solution of the inhomogeneous Volterra equation (\ref{eq:Bcondition}) under the premises that $G^{(i)}(\chi)$ is twice continuously differentiable, $G^{(i)}(\chi_{\rm hor})=0$ and $\dd G^{(i)}/ \dd \chi |_{\chi_{\rm hor}} =0$. If one specifies $G^{(i)}(\chi)$ down to a value $\chi_{\rm min}$, then $B^{(i)}(\chi)$ is well-defined in the range $\bb{\chi_{\rm min},\chi_{\rm hor}}$ by (\ref{eq:Bcondition}). Note that if we dropped the assumption of a flat universe, (\ref{eq:Bcondition}) would still be solvable, but analytical progress would be hampered.

To find the solution of the homogeneous equation, obtained from (\ref{eq:Bcondition}) by setting $G^{(i)}(\chi) \equiv 0$, we define 
\eq{
\label{eq:defforhom}
b(\chi) \equiv \frac{B^{(i)}(\chi)}{\chi}\; H(\chi_{\rm hor}-\chi)\;; ~~~f(\chi) \equiv \chi H(\chi)\;,
}
where $H(\chi)$ denotes the Heaviside step function. Then (\ref{eq:Bcondition}) can be re-written as a cross-correlation,
\eq{
G^{(i)}(\chi) = \int_{-\infty}^\infty \dd \bar{\chi}\; b(\bar{\chi})\; f(\bar{\chi}-\chi) = \bc{b * f} (\chi)\;.
}
The introduction of the Heaviside functions in (\ref{eq:defforhom}) was used to extend the integration to zero and infinity. If we denote Fourier transforms by a tilde, the convolution theorem yields $\tilde{G}^{(i)} = \tilde{b}\, \tilde{f}$. From this equation is it readily seen that for $G^{(i)}(\chi) \equiv 0$ it follows $B^{(i)}(\chi) \equiv 0$ in the interval $\bb{\chi,\chi_{\rm hor}}$. Hence the solution of the homogeneous Volterra equation consists only of the trivial one and (\ref{eq:volterrasolution}) constitutes the full, unique solution of (\ref{eq:Bcondition}). In summary, for a given $G^{(i)}(\chi)$ that fulfils the conditions imposed by (\ref{eq:modGI}) and (\ref{eq:modGG}), we can calculate the corresponding weight function via (\ref{eq:volterrasolution}) and use the result to construct transformed power spectra (\ref{eq:defpi}).

Note the analogy between (\ref{eq:Bcondition}) and the definition of the lensing efficiency (\ref{eq:lenseff}). This can be interpreted as $G^{(i)}(\chi)$ being a modified lensing efficiency, which is then used to construct an alternative lensing convergence with desired properties chosen via $G^{(i)}(\chi)$. For details on this view see the motivation of the nulling technique given in \citet{joachimi08b}.

\section{Construction of weights}
\label{sec:weights}

Apart from the requirements formulated in Sect.$\,$\ref{sec:trafo} to ensure a boosting of the GI signal with respect to cosmic shear, the choice of $G^{(i)}(\chi)$ is arbitrary. In the following we choose a specific parametrisation of $G^{(i)}(\chi)$ which is convenient and intuitive, but not necessarily optimal. Its base is a Gaussian that is peaked at $\chi_i$, which fosters a strong contribution of GI correlations via the first term of (\ref{eq:modGI}). Some additional flexibility is needed at $\chi < \chi_i$, allowing for sign changes of $G^{(i)}(\chi)$ to downweight the lensing signal. We define
\eq{
\label{eq:defG}
G^{(i)}(\chi) \equiv {\cal N}\; \exp \bc{-\frac{(\chi-\chi_{\rm m})^2}{\sigma^2}} \br{\chi - b}\;,
}
where ${\cal N}$, $\sigma$, $b$, and $\chi_{\rm m}$ are free parameters. All four parameters depend on the choice of galaxy sample $i$, but we do not specify this dependence for reasons of better readability. The first derivative of $G^{(i)}(\chi)$ with respect to comoving distance reads
\eq{
\label{eq:Gderivative}
\frac{\partial G^{(i)}}{\partial \chi} (\chi) = {\cal N}\; \exp \bc{-\frac{(\chi-\chi_{\rm m})^2}{\sigma^2}} \bc{1 - 2\, (\chi - b)\; \frac{\chi-\chi_{\rm m}}{\sigma^2}}\;.
}
From this result and by means of (\ref{eq:volterrasolution}) one readily obtains the weight function
\eqa{
B^{(i)}(\chi) &=& {\cal N}\; \frac{2 \chi}{\sigma^2}\; \exp \bc{-\frac{(\chi-\chi_{\rm m})^2}{\sigma^2}}\\ \nn
&& \hspace*{1.cm} \times\; \bc{2\, (\chi - b)\; \frac{(\chi-\chi_{\rm m})^2}{\sigma^2} - 3 \chi + 2 \chi_{\rm m} + b}\;.
}
The normalisation of $G^{(i)}(\chi)$ is related to the one of $B^{(i)}(\chi)$ via (\ref{eq:Bcondition}), but is otherwise irrelevant to the problem. We fix ${\cal N}$ by requiring
\eq{
\label{eq:normalisation}
\int_{\chi_{\rm min}}^{\chi_{\rm hor}} \dd \chi\; \bc {B^{(i)}(\chi) }^2 \approx \sum_{j=j_{\rm min}}^{N_z} \bc {B^{(i)}(\chi(z_j)) }^2 \chi'(z_j)\; \Delta z_j = 1\;.
}
Note that since ${\cal N}$ depends on the other free parameters, e.g. $\sigma$, a consistent normalisation is actually important when studying $G^{(i)}(\chi)$ as a function of these parameters, as we will do in Sects.$\,$\ref{sec:performance} and \ref{sec:nulling}.

\begin{figure}[t]
\centering
\includegraphics[scale=0.365,angle=270]{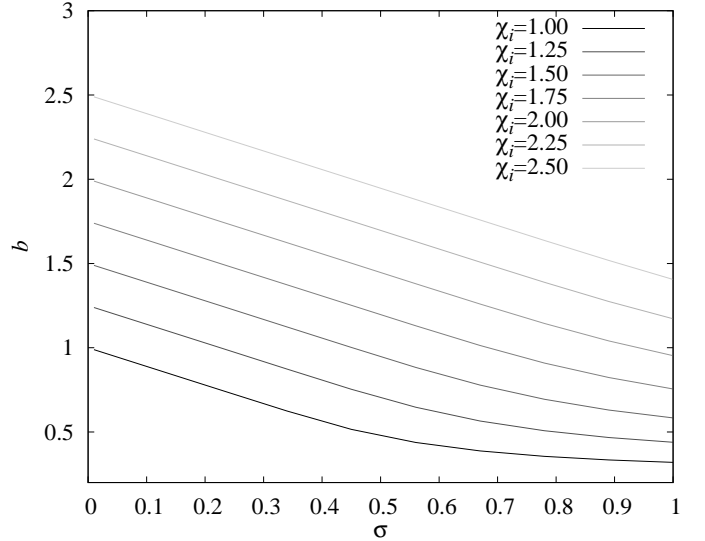}
\caption{Parameter $b$ as a function of $\sigma$, i.e. the width of the Gaussian in $G^{(i)}(\chi)$. Plotted are the results for different peak positions $\chi_i$ as indicated in the legend. For most of the considered range, (\ref{eq:bapprox}) provides an excellent approximation. Only for small $\chi_i$ in combination with large $\sigma$ do the curves start to level off. Note that $b$, $\sigma$, and $\chi_i$ are given in abstract units of 1 in this plot.}
\label{fig:bnumeric}
\end{figure}

\begin{figure}[t]
\centering
\includegraphics[scale=.36,angle=270]{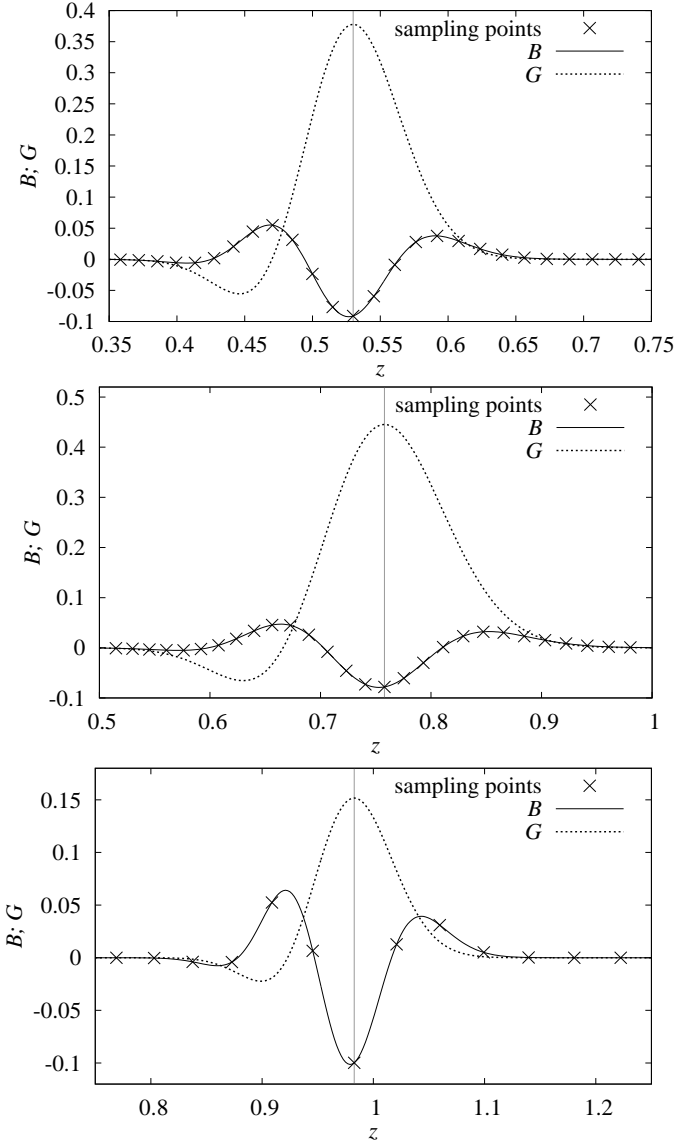}
\caption{Functions $G^{(i)}(\chi(z))$ and $B^{(i)}(\chi(z))$ for different $z_i$, as indicated by the vertical grey lines, and redshift binnings, using the width $\sigma_z$ determined via (\ref{eq:diagnostic}) in each case. In addition the sampling points corresponding to the median redshifts of the bins are shown for $B^{(i)}(\chi(z))$. The normalisation has been chosen according to (\ref{eq:normalisation}). See Table \ref{tab:configs} for an overview on the survey models referred to in the following. \textit{Top panel}: For the spectroscopic survey S and $z_i=0.53$; $\sigma_z=0.055$. \textit{Centre panel}: For the survey with good photometric redshifts P1 and $z_i=0.76$; $\sigma_z=0.085$. \textit{Bottom panel}: For the survey with standard photometric redshifts P2 and $z_i=0.98$; $\sigma_z=0.055$.}
\label{fig:BG}
\end{figure}

Two of the remaining three free parameters of (\ref{eq:defG}) will now be used to boost (\ref{eq:modGI}) and suppress (\ref{eq:modGG}). First, we demand that (\ref{eq:defG}) is peaked at $\chi_i$, i.e. $\partial G^{(i)}/ \partial \chi\, |_{\chi_i} = 0$. Using (\ref{eq:Gderivative}), we obtain
\eq{
\label{eq:conditionpeak}
\chi_{\rm m} = \chi_i - \frac{\sigma^2}{2} \br{\chi_i - b}^{-1}\;.
}
The second condition should render the integral in (\ref{eq:modGG}) close to zero. While it is possible to numerically determine for instance the parameter $b$ such that this condition is fulfilled for every angular frequency individually, we prefer to proceed in a way that does not rely on a model of cosmic shear power spectra at all. We note that if the width of the Gaussian $\sigma$ is relatively small, the support of the integral in (\ref{eq:modGG}) has a small range and hence $P_\delta$ can be well approximated as only varying slowly. The dependence on redshift should be roughly $P_\delta(k,z) \propto D(z)^2 \propto (1+z)^{-2}$, where $D(z)$ is the linear growth factor for which we assumed $D(z) \propto (1+z)^{-1}$ as holds true in the matter-dominated epoch. This redshift dependence then cancels the $(1+z)^2$ term in (\ref{eq:modGG}), so that we consider the condition
\eq{
\label{eq:conditionGint}
\int_0^{\chi_i} \dd \chi \br{1 - \frac{\chi}{\chi_i}} G^{(i)}(\chi) = 0\;.
}
Inserting (\ref{eq:defG}) together with (\ref{eq:conditionpeak}), and making the further definitions $y \equiv (\chi-\chi_i)/\sigma$ and $m \equiv (\chi_i- b)/\sigma$, we transform this integral as follows,
\eqa{
\label{eq:proxyforb}
&& \int_0^{\chi_i} \dd \chi \br{1 - \frac{\chi}{\chi_i}} G^{(i)}(\chi)\\ \nn
&\approx& - \frac{{\cal N}\, \sigma^3}{\chi_i} \int_{-\infty}^0 \dd y\; y\, (y+m)\; \exp \bc{- \br{ y + \frac{1}{2m} }^2 }\\ \nn
&=& - \frac{{\cal N}\, \sigma^3}{8 m^2 \chi_i} \bc{ 2m\, (1-2m^2) \expo{-1/(4m)^2} + \sqrt{\pi} \bb{1+{\rm Erf}\br{\frac{1}{2m}}} }\;.
}
The approximation in the first equality refers to replacing the lower boundary of the integral $-\chi_i/\sigma$ by $-\infty$, which is valid if $\chi_i/\sigma \gg 1$, i.e. if $G^{(i)}(\chi)$ is compact ($\sigma \ll 1$) and peaks not too close to zero ($\chi_i \gg 0$). The root of the term in curly brackets can be found numerically, resulting in
\eq{
\label{eq:bapprox}
b(\sigma,\chi_i) = \chi_i - m\, \sigma  ~~~\mbox{with}~~ m \approx 1.10687\;.
}
We have solved (\ref{eq:conditionGint}) directly and plot the resulting $b$ in Fig.$\,$\ref{fig:bnumeric}. We find excellent agreement with the approximate solution (\ref{eq:bapprox}) as long as the assumption discussed above is fulfilled. Significant deviations from the linear behaviour of $b$ as a function of $\sigma$ are only found for $\chi_i/\sigma \lesssim 2$. For reasons of simplicity we will restrict ourselves to cases where the approximation (\ref{eq:bapprox}) holds. This means in particular that we will not consider signals at very small redshifts, where $\chi_i$ is necessarily small. In practice we use the condition $G^{(i)}(\chi(z_{\rm min})) \approx 0$ with $z_{\rm min}$ the minimum redshift used in the survey as a simple cross-check to ensure that this approximation is sufficiently accurate.

The conditions specified above are strictly fulfilled only for continuous $\chi$ or $z$. However, we will in practice use the discretised transformation (\ref{eq:defpidiscrete}) and thus have to make sure that GI boosting and GG suppression work accurately also in this case. Via a procedure outlined in the following, we optimise the remaining free parameter $\sigma$ to guarantee a good sampling of $G^{(i)}(\chi)$ by the discrete set of weights $B^{(i)}(\chi(z_j))$ with $j=1,\,..\,,N_z$, thereby fulfilling $\partial G^{(i)}/ \partial \chi\, |_{\chi_i} = 0$ and (\ref{eq:conditionGint}) to good accuracy.

As the sampling points of (\ref{eq:defpidiscrete}) we choose the medians of the redshift distributions of the galaxy samples employed. It is expected that the optimal choice of the parameter $\sigma$, denoted by $\sigma_{\rm opt}$ in the following, will depend intricately on the positions of these sampling points and hence on the redshift distributions of the different galaxy samples in the cosmic shear data, in particular if the number of sampling points is small, e.g. if the distributions have a large scatter. Since the binning is done in terms of redshift, it is convenient to work with the quantity $\sigma_z \equiv \bc{ \chi'(z_i) }^{-1} \sigma$ instead of $\sigma$. We will also give our choices of $\sigma_{\rm opt}$ in terms of $\sigma_z$ throughout.

We introduce the discrete version of the function $G^{(i)}(\chi)$,
\eq{
\label{eq:defgdiscrete}
{G'}^{(i)} (\chi(z_k)) \equiv \sum_{j=k}^{N_z} B^{(i)}(\chi(z_j))\; \br{1-\frac{\chi_(z_k)}{\chi(z_j)}}\; \chi'(z_j)\; \Delta z_j\;.
}
Then we consider the root mean square deviation of all function values ${G'}^{(i)} (\chi(z_k))$ used,
\eq{
\label{eq:diagnostic}
\zeta(\sigma_z) \equiv \sqrt{ \frac{1}{N_z} \sum_{k=1}^{N_z} \left| {G'}^{(i)} (\chi(z_k),\sigma_z) - G^{(i)} (\chi(z_k),\sigma_z) \right|^2 }\;,
}
as a criterion for how well $G^{(i)}(\chi)$ is sampled by the discrete set of function values $B^{(i)}(\chi(z_j))$ entering (\ref{eq:defgdiscrete}). In the equation above we have made the dependence on $\sigma_z$ explicit in the arguments. We emphasise that the determination of $\sigma_z$ via the diagnostic $\zeta$ is optimal only in the sense that it allows us to find a representative sampling of $G^{(i)}(\chi)$ such that $\partial G^{(i)}/ \partial \chi\, |_{\chi_i} = 0$ and (\ref{eq:conditionGint}) hold to good accuracy. It will in general not yield an optimal amplification of the GI signal over the lensing signal, which depends on the explicit form of both signals. 

In Fig.$\,$\ref{fig:BG} we have plotted a selection of typical results for $G^{(i)}(\chi)$ and the corresponding weight function $B^{(i)}(\chi)$. As common features of $G^{(i)}(\chi)$ a distinct peak at $z_i$ and a negative dip at $z<z_i$, the latter necessary to fulfil (\ref{eq:conditionGint}), are discernible. The weight function $B^{(i)}(\chi)$ has three pronounced extrema of which the central one is located at $z_i$, plus a shallow fourth one at low redshift. 

Note that the method laid out here is completely independent of any assumptions about the angular dependence of both the underlying lensing and intrinsic alignment signals. To determine the weights entering (\ref{eq:defpidiscrete}), we only make use of the well-known redshift dependence of the GI and GG signals, plus the redshift binning of the survey to be analysed. We note that the weights $B^{(i)}(\chi(z_j))$ depend on $\Omega_{\rm m}$ and possibly further cosmological parameters via the distance-redshift relation. However, the same applies to the weights used in the standard nulling technique, and from the investigation by \citet{joachimi09} we conclude that this dependence is weak and that the assumption of an incorrect cosmology when constructing the weights is uncritical.

\section{Modelling}
\label{sec:modelling}

To assess the performance of the boosting technique, we need to model both the cosmic shear and the intrinsic alignment signals. To this end, we assume a spatially flat $\Lambda$CDM universe with matter density parameter $\Omega_{\rm m}=0.25$ and Hubble parameter $h=0.7$. The matter power spectrum has a primordial slope $n_{\rm s}=1.0$ and normalisation $\sigma_8=0.8$. The transfer function is computed according to \citet{eisenstein98}, using a baryon density parameter of $\Omega_{\rm b}=0.05$, while the non-linear evolution of the power spectrum is determined by the fit formula of \citet{smith03}.

We use the linear alignment model \citep{catelan01,hirata04} to calculate the matter density-intrinsic power spectrum,
\eq{
\label{eq:GIlinalign}
P_{\delta {\rm I}}\br{k,z} = - C_{\rm GI}\; \rho_{\rm cr}\; \frac{\Omega_{\rm m}\;(1+z)^2}{D(z)}\; P_{\delta}\br{k,z}\;,
}
where the normalisation is chosen such that $C_{\rm GI}\, \rho_{\rm cr} \approx 0.0134$ (\citealp{joachimi09}, and references therein) with $\rho_{\rm cr}$ the critical density. In (\ref{eq:GIlinalign}) we use the full non-linear matter power spectrum as suggested by \citet{bridle07}. While this conjecture lacks a sound physical basis, the resulting signal fits existing data well \citep{bridle07} and has recently been shown to also be consistent with halo model calculations \citep{schneiderm09}.

A cosmic shear survey is modelled by assuming an overall galaxy redshift distribution according to \citet{smail94},
\eq{
\label{eq:redshiftdistribution}
p_{\rm tot}(z) \propto \br{\frac{z}{z_0}}^2 \exp \bc{ -\br{\frac{z}{z_0}}^\beta}
}
with $z_0=0.64$ and $\beta=1.5$, corresponding to a median redshift of $z_{\rm med}=0.9$. We cut the distribution below $z_{\rm min}=0.2$ and above $z_{\rm max}=2.0$ and normalise (\ref{eq:redshiftdistribution}) in that interval. The overall redshift distribution is then sliced into disjoint bins. In those cases where a scatter due to photometric redshift estimates is present, we assume the distribution of photometric redshifts for a given true redshift to be a Gaussian, centred on the true redshift and with a width of $\sigma_{\rm ph}(1+z)$. The distributions of true redshifts $p^{(i)}(z)$ for each photometric redshift bin $i$ are then computed according to a scheme detailed in \citet{joachimi09}.

We consider three different survey models which are summarised in Table \ref{tab:configs}. All of these surveys are assumed to cover the whole extragalactic sky, i.e. $A_{\rm survey}=20,000\,{\rm deg}^2$. To calculate shape noise, we use an intrinsic ellipticity dispersion of $\sigma_\epsilon=0.35$ throughout. 

First, we construct a \lq spectroscopic\rq\ survey S for which redshift bins are assigned with width $0.01(1+z)$ and no scatter. In this case the signals are calculated to excellent approximation not over the complete bin width, but at the median redshifts of each bin. Whilst it is in principle possible to achieve such a dense redshift binning and small scatter with photometric redshifts \citep[see e.g.][]{ilbert08}, it is more likely that future large-area spectroscopic surveys fit into this category. In any case the number of available galaxies will be small. Taking the wide spectroscopic survey of the Euclid mission as reference \citep{laureijs09}, we set the overall galaxy number density to $n_{\rm g}=1\,{\rm arcmin}^{-2}$.

\begin{table}[t]
\begin{minipage}[t]{\columnwidth}
\centering
\caption{Overview on the different survey models used.}
\begin{tabular}[t]{ccccc}
identifier & redshifts & bin width & $\sigma_{\rm ph}$ & $n_{\rm g}\,[{\rm arcmin}^{-2}]$\\
\hline\hline
S  & spectroscopic    & $0.01(1+z)$ & 0    & 1\\
P1 & good photo-z     & $0.01(1+z)$ & 0.03 & 10\\
P2 & standard photo-z & $0.02(1+z)$ & 0.05 & 40\\
\end{tabular}
\label{tab:configs}
\end{minipage}
\end{table}

Second, we create a survey that features high-quality photometric redshift data, termed P1. We choose the same binning scheme as for the first case, but introduce a photometric redshift scatter of $\sigma_{\rm ph}=0.03$, corresponding to the target value of the Euclid imaging survey. To be conservative, we assume that this photometric redshift quality is only attainable for a subset of galaxies and set $n_{\rm g}=10\,{\rm arcmin}^{-2}$. Finally, we make use of a setup P2 with redshift binning in steps of $0.02(1+z)$ and scatter $\sigma_{\rm ph}=0.05$, which can be regarded as representative of a standard future imaging survey designed to do cosmic shear. Again referring to \citet{laureijs09}, we adopt $n_{\rm g}=40\,{\rm arcmin}^{-2}$ in this case.

The photometric redshift bin widths are chosen such that the associated distributions of neighbouring bins can still be well distinguished. We have found that narrowing the bin widths substantially below about $1/3\, \sigma_{\rm ph}$ deteriorates the performance of the boosting technique. It should be noted that spectroscopic redshifts as well as photometric redshifts of high quality are usually limited to a brighter subset of galaxies, therefore altering the overall redshift distribution of galaxies. However, to facilitate the comparison between the three survey models under scrutiny, we keep $p_{\rm tot}(z)$ as specified above.

With the three-dimensional GG and GI power spectra and the redshift distributions $p^{(i)}(\chi) = p^{(i)}(z) / \chi'(z)$ at hand, one can calculate the tomographic power spectra according to (\ref{eq:GGdef}) and (\ref{eq:GIdef}). For the further analysis we divide the angular frequency range into $N_\ell=200$ logarithmic bins between $\ell=10$ and $\ell=20000$.

\section{Performance of GI boosting}
\label{sec:performance}

\subsection{Boosted signals}
\label{sec:boostedsignals}

\begin{figure}[t]
\centering
\includegraphics[scale=.35,angle=270]{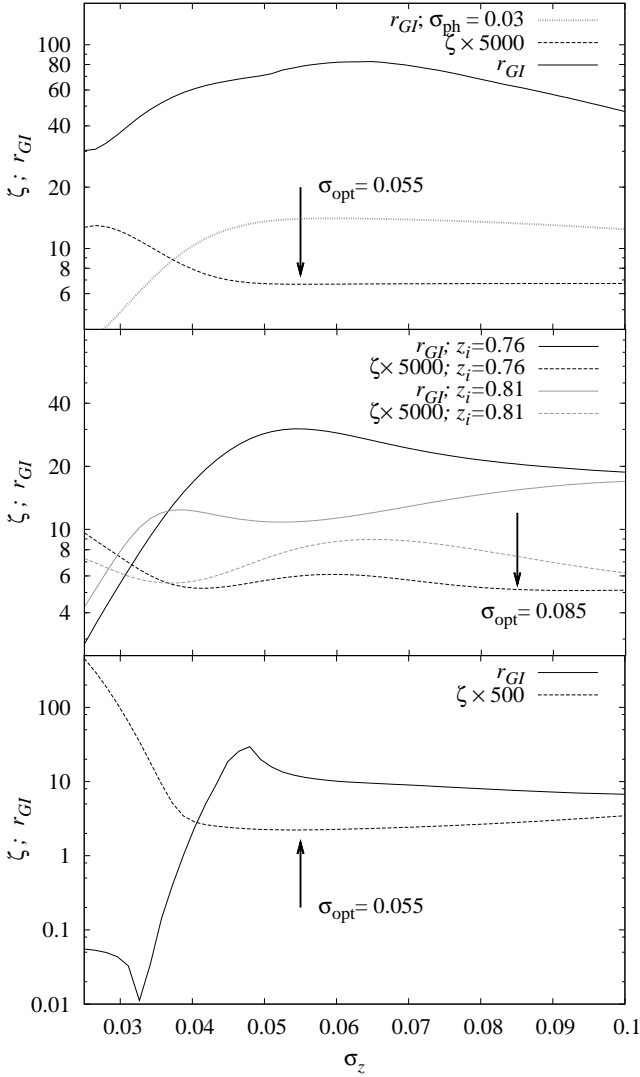}
\caption{Diagnostic $\zeta$, see (\ref{eq:diagnostic}), and GI over GG ratio $r_{\rm GI}$, see (\ref{eq:rgi}), as a function of $\sigma_z$. Shown is $r_{\rm GI}$ as solid curve and $\zeta$ as dashed curve. Since the normalisation of $\zeta$ is arbitrary, we have rescaled $\zeta$ for easier inspection. In each panel the choice of $\sigma_{\rm opt}$ is marked with an arrow. Note that this choice was made without resorting to $r_{\rm GI}$ which is not measurable from real data. \textit{Upper panel}: For the spectroscopic survey S at $z_i=0.53$. In addition we have plotted $r_{\rm GI}$ for the case $\sigma_{\rm ph} = 0.03$. Note that $\zeta$ remains the same in both cases. \textit{Centre panel}: For the survey P1 ($\sigma_{\rm ph} = 0.03$) at $z_i=0.76$. Note the dip of $\zeta$ at $\sigma_z \sim 0.04$ which is caused by the weight function $B^{(i)}(\chi)$ being sampled close to its extrema. We have added the curves for $z_i=0.81$ as grey lines, where $B^{(i)}(\chi)$ is sampled almost exactly at its extrema, leading to a local maximum in $r_{\rm GI}$ at $\sigma_z = 0.038$, traced well by a corresponding minimum in $\zeta$. \textit{Lower panel}: For the survey P2 ($\sigma_{\rm ph}=0.05$) at $z_i=0.98$.}
\label{fig:sigopt}
\end{figure}

To condense the performance of the boosting technique into a single number, we define the median with respect to angular frequency of the ratio of GI over GG signal,
\eq{
\label{eq:rgi}
r_{\rm GI} \equiv {\rm median}\bc{\left| \frac{X_{\rm GI}(\ell_i)}{X_{\rm GG}(\ell_i)} \right|}_{i=1,\,..\,,N_\ell}\;,
}
where $X$ can be replaced by any tomography power spectrum $P^{(ij)}(\ell)$ or the transformed power spectra $\Pi^{(i)}(\ell)$. Note that this quantity is not available from a real survey because we are not able to separate the GG and GI signals, but only extract their sum from the data. We have chosen the median in (\ref{eq:rgi}) since we find that the mean is not a robust measure for two reasons. First, if the GG signal is suppressed by several orders of magnitude, numerical noise stemming from the computation of the power spectra can become important, leading to unphysical dips in the residual power spectrum. Second, the residual GG signal may have sign changes close to which $r_{\rm GI}$ becomes very large, thus dominating the mean. Both effects would mimic a stronger boosting than is actually observed.

In Fig.$\,$\ref{fig:sigopt} we show $r_{\rm GI}$, together with the diagnostic $\zeta$ as defined in (\ref{eq:diagnostic}), as a function of $\sigma_z$ for one $z_i$ per survey model. Overall we find that small values of $\zeta$ indeed indicate regimes of $\sigma_z$ in which the GI signal is well boosted. It is important to note that the absolute value of $\zeta$ is meaningless due to the arbitrariness in the overall amplitude of $G^{(i)}(\chi)$. When $G^{(i)}(\chi)$ is no longer well sampled for small $\sigma_z$, $\zeta$ features a clear increase. Sometimes secondary minima in $\zeta$ can be observed, see the centre panel of Fig.$\,$\ref{fig:sigopt}, which is caused by the sampling points being consecutively placed at the extrema of $B^{(i)}(\chi)$. Thereby, although only sparsely sampled, the discrete form (\ref{eq:defgdiscrete}) captures the main characteristics of $B^{(i)}(\chi)$ and hence can well represent $G^{(i)}(\chi)$, yielding a small value of $\zeta$.

\begin{table}[t]
\begin{minipage}[t]{\columnwidth}
\centering
\caption{Summary of $r_{\rm GI}$ for different values of $z_i$ and the three survey models. Given are values of $r_{\rm GI}$ of the original power spectra for the auto-correlation (\lq auto\rq), the cross-correlation between bin $i$ and the background bin with index $(i+N_z)/2$ (\lq mid\rq), and the cross-correlation between bin $i$ and the most distant bin with index $N_z$ at $z \lesssim 2$ (\lq far\rq). The tag \lq boost\rq\ stands for the transformed signals. In addition $\sigma_{\rm opt}$ is listed for every considered case.}
\begin{tabular}[t]{ccccccc}
$\!\!\!$survey$\!\!\!$ & $z_i$ & $\sigma_{\rm opt}$ &$r_{\rm GI}$(auto) & $\!r_{\rm GI}$(mid)$\!$ & $\!r_{\rm GI}$(far)$\!$ & $r_{\rm GI}$(boost)\\
\hline\hline
S  & 0.53 & 0.055 & 0.04 & 0.47 & 0.56 & 78.26\\
   & 0.76 & 0.085 & 0.02 & 0.25 & 0.32 & 415.79\\
   & 0.96 & 0.095 & 0.01 & 0.16 & 0.22 & 62.10\\
P1 & 0.53 & 0.055 & 0.12 & 0.46 & 0.54 & 13.56\\
   & 0.76 & 0.085 & 0.06 & 0.24 & 0.31 & 20.49\\
   & 0.96 & 0.095 & 0.04 & 0.16 & 0.21 & 25.86\\
P2 & 0.52 & 0.045 & 0.19 & 0.47 & 0.55 & 5.97\\
   & 0.74 & 0.050 & 0.10 & 0.26 & 0.32 & 7.75\\
   & 0.98 & 0.055 & 0.06 & 0.15 & 0.20 & 11.63\\
\end{tabular}
\label{tab:rgi}
\end{minipage}
\end{table}

In the top panel of Fig.$\,$\ref{fig:sigopt} $r_{\rm GI}$ for both surveys S and P1 is given. Since the binning scheme is identical for both surveys, $\zeta$ is the same. This example demonstrates that $r_{\rm GI}$ depends considerably on the details of the actual signals, in this case a change from $\sigma_{\rm ph} = 0$ to $\sigma_{\rm ph} = 0.03$. The diagnostic $\zeta$ does not trace the boosting of the actual signals and can consequently not be exploited to find the maximum $r_{\rm GI}$. However, for both surveys $\zeta$ identifies the regime of small $\sigma_z$ in which the boosting performs worse and which thus should be avoided. In the case $\sigma_{\rm ph} = 0.05$ the sampling in redshift becomes fully insufficient for small $\sigma_z$. Accordingly, $\zeta$ rises sharply, and the GG signal starts to dominate again.

The optimal width of $G^{(i)}(\chi)$ can be chosen freely in the interval where $\zeta$ is stable and small. If there is a clear minimum, we place $\sigma_{\rm opt}$ there; otherwise we set $\sigma_{\rm opt}$ to a small value in the interval where $\zeta$ is small, see e.g. the centre panel of Fig.$\,$\ref{fig:sigopt}. This assignment of $\sigma_{\rm opt}$ may not be unique, but it is uncritical. Note that the weight functions corresponding to the optimum cases of the examples shown in Fig.$\,$\ref{fig:sigopt} are those depicted in Fig.$\,$\ref{fig:BG}. We emphasise again that $r_{\rm GI}$ cannot be measured from real data, and accordingly we do not use this quantity to determine $\sigma_{\rm opt}$.

One might expect that the denser the sampling points of $G^{(i)}(\chi)$ and $B^{(i)}(\chi)$ can be placed, the more sharply peaked weight functions can be well represented by the discrete sampling, and thus smaller values of $\sigma_z$ could be chosen. However, consider the case $\sigma_{\rm ph} = 0.05$ and $z_i=0.98$ which is shown in the bottom panels of both Figs.$\,$\ref{fig:BG} and \ref{fig:sigopt}. Although $\sigma_{\rm opt}$ is small compared to e.g. our findings for survey P1, the sparse sampling obviously captures the main features of the weight function and hence results in a small $\zeta$.

\begin{figure*}[t]
\centering
\includegraphics[scale=.35,angle=270]{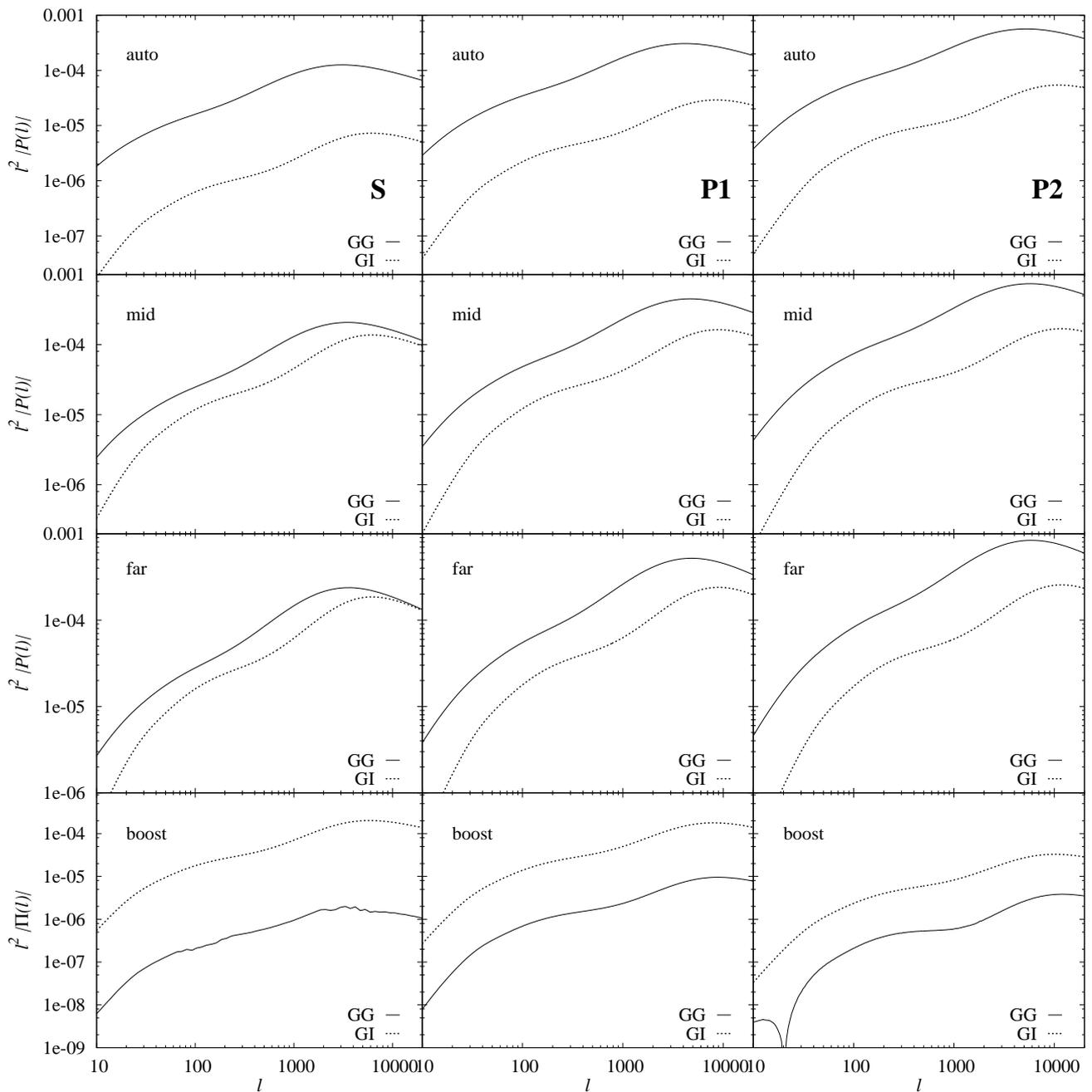}
\caption{\textit{Left column}: Example set of lensing (GG) and intrinsic alignment (GI) tomography power spectra for the spectroscopic survey S and $z_i=0.53$. The GG signal is shown as solid line, the GI signal as dotted line. The upper three panels show power spectra for different background bins $j$, i.e. auto-correlations ($j=i$, \lq auto\rq), cross-correlations with a bin at intermediate redshift ($j=(i+N_z)/2$, \lq mid\rq), and cross-correlations with the most distant bin ($j=N_z$, \lq far\rq). In the bottom panel the transformed GG and GI signals are plotted (\lq boost\rq). Note that absolute values of the power spectra are shown throughout. \textit{Centre column}: Same as above, but for the survey P1 ($\sigma_{\rm ph}=0.03$) and $z_i=0.76$. \textit{Right column}: Same as above, but for the survey P2 ($\sigma_{\rm ph}=0.05$) and $z_i=0.98$.}
\label{fig:ps}
\end{figure*}

In Fig.$\,$\ref{fig:ps} we have plotted example sets of original and transformed power spectra, the latter each computed for the optimal values of $\sigma_z$. Table \ref{tab:rgi} lists the corresponding values of $r_{\rm GI}$ and $\sigma_{\rm opt}$ for the three survey models and three redshifts $z_i$ each, including the cases depicted in the figure. The GI over GG ratio $r_{\rm GI}$ for the original power spectra ranges from about $1\,\%$ to $50\,\%$. For a correlation between galaxy samples $i$ and $j$ with $z_i < z_j$, $r_{\rm GI}$ increases strongly with the separation between $z_j$ and $z_i$. Both the GI and GG signals show this behaviour due to (\ref{eq:lenseff}). Since the cosmic shear signal is generated by all the matter between $z=0$ and $z_i$ with the highest efficiency at $z_i/2$, whereas the intrinsic alignment contribution stems form matter around $z_i$, the GI signal has the stronger dependence on redshift, causing the increase in $r_{\rm GI}$. For the non-linear version of the linear alignment model this effect can lead to a GI signal whose absolute value can come close to or even surpass the cosmic shear signal for large $z_j-z_i$, see e.g. also \citet{bridle07}.

It is evident from Table \ref{tab:rgi} that the better resolved the redshift information is, the more can the GI signal be boosted. For quasi-spectroscopic data, the residual GG contribution is well below the $2\,\%$-level and hence expected to be negligible. In the case $z_i=0.76$ we find by chance a near-total cancellation of the cosmic shear signal. For good photo-z data with $\sigma_{\rm ph} = 0.03$ the method is also effective, yielding $r_{\rm GI}$ well in excess of 10, so that any biases due to the residual GG contribution are likely to remain below the statistical errors of intrinsic alignment parameters. For survey P2 it is still possible to produce a dominating GI signal, with $r_{\rm GI}$ between approximately 6 and 12, but a GG residual exceeding $10\,\%$ may require further treatment to avoid a significant bias.

\subsection{Parameter constraints}
\label{sec:constraints}

The boosted GI signal has the potential use of directly constraining models of intrinsic alignments, provided that the statistical power is sufficiently high and that systematics due to residual GG contributions are under control. We set up a simple intrinsic alignment model and use the Fisher matrix formalism \citep{tegmark97} to forecast expected errors and biases on its free parameters. We define
\eq{
\label{eq:GImodel}
P^{\rm model}_{\delta {\rm I}}\br{k,z} = A\; P_{\delta {\rm I}}\br{k,z}\; \br{\frac{1+z}{1+z_{\rm piv}}}^\gamma\;,
}
where $P_{\delta {\rm I}}$ is given by the linear alignment model (\ref{eq:GIlinalign}). The free parameters are $A$ and $\gamma$, i.e. we allow for an arbitrary signal amplitude and an additional redshift dependence. The fiducial model is (\ref{eq:GIlinalign}), so $A=1$ and $\gamma=0$, and we set $z_{\rm piv}=0.3$. The same parametrisation was e.g. used by \citet{mandelbaum09}.

\begin{table}[t]
\begin{minipage}[t]{\columnwidth}
\centering
\caption{Statistical errors $\sigma_{\rm stat}$ and residual biases $b_{\rm sys}$ for the different survey models used. The left column shows marginalised $1\,\sigma$ errors on the amplitude $A$ of the GI signal and on $\gamma$, quantifying an additional redshift dependence. In the right column results obtained for only varying $A$ are listed.}
\begin{tabular}[t]{cc|cc|cc}
survey & parameter & \multicolumn{2}{|c|}{$\gamma$ varied} & \multicolumn{2}{|c}{$\gamma$ fixed}\\
 & & $\sigma_{\rm stat}$ & $b_{\rm sys}$ &  $\sigma_{\rm stat}$ & $b_{\rm sys}$\\
\hline\hline
S  & $A$      & 2.885 & -0.009 & 0.827 & -0.004\\
   & $\gamma$ & 7.356 &  0.012 &       &       \\
P1 & $A$      & 0.712 & -0.081 & 0.172 & -0.046\\
   & $\gamma$ & 1.776 &  0.090 &       &       \\
P2 & $A$      & 0.697 & -0.181 & 0.272 & -0.171\\
   & $\gamma$ & 2.017 &  0.031 &       &       \\
\end{tabular}
\label{tab:boost_fisher}
\end{minipage}
\end{table}

Assuming that the signal covariance is itself not parameter-dependent, which holds to very good accuracy for the large survey we consider \citep[][]{eifler08}, the Fisher matrix reads
\eq{
\label{eq:fisher}
F_{\mu \nu} = \sum_\ell \sum_{i,j} \frac{\partial \Pi_{\rm GI}^{(i)}(\ell)}{\partial p_\mu}\; {\rm Cov}^{-1}\br{\Pi_{\rm GI}^{(i)}(\ell),\Pi_{\rm GI}^{(j)}(\ell)}\; \frac{\partial \Pi_{\rm GI}^{(j)}(\ell)}{\partial p_\nu}\;,
}
for a parameter vector $\vek{p}=\bc{A,\gamma}$. Using (\ref{eq:defpidiscrete}), one can readily relate the covariance of the transformed power spectra to that of the original power spectra,
\eqa{
{\rm Cov}\br{\Pi_{\rm GI}^{(i)}(\ell),\Pi_{\rm GI}^{(j)}(\ell)} &=& \sum_{k,l=0}^{N_z} B^{(i)}(\chi(z_k))\; B^{(j)}(\chi(z_l))\\ \nn
&& \hspace*{-1cm} \times\; {\rm Cov}\br{P_{\rm GI}^{(ik)}(\ell),\; P_{\rm GI}^{(jl)}(\ell)}\; \chi'(z_k)\; \chi'(z_l)\; \Delta z_k\; \Delta z_l\;.
}
The power spectrum covariance in turn is given by \citep[see][and references therein]{joachimi08}
\eqa{
\label{eq:covPtom}
{\rm Cov}\br{P_{\rm GI}^{(ij)}(\ell),\; P_{\rm GI}^{(kl)}(\ell)} &=& \frac{2\pi}{A_{\rm survey} \ell \Delta \ell}\; \biggl( \bar{P}_{\rm GI}^{(ik)}(\ell)\bar{P}_{\rm GI}^{(jl)}(\ell) \\ \nn
&& \hspace*{-2cm} +\; \bar{P}_{\rm GI}^{(il)}(\ell)\bar{P}_{\rm GI}^{(jk)}(\ell) \biggr) ~~~~~\mbox{with}~~  \bar{P}_{\rm GI}^{(ij)} = P_{\rm GI}^{(ij)}+\delta_{ij} \frac{\sigma_\epsilon^2}{2\bar{n}^{(i)}}\;,
}
where $\Delta \ell$ is the width of the angular frequency bins and $\bar{n}^{(i)}$ the number of galaxies belonging to sample $i$. Equation (\ref{eq:covPtom}) holds under the assumptions of Gaussian density fluctuations, a uniform sampling of galaxies, and a simple survey geometry where the scales considered are much smaller than the extent of the survey.

The bias formalism \citep[e.g.][]{huterer06,amara08,joachimi09} allows us to compute the bias on the intrinsic alignment parameters due to the residual GG signal in the transformed power spectra via
\eqa{
\label{eq:bias}
b_{\rm sys}(p_\mu) &=& \sum_\nu \br{ F^{-1}}_{\mu \nu}\\ \nn
&& \hspace*{0cm} \times\; \sum_\ell \sum_{i,j} \Pi_{\rm GG}^{(i)}(\ell)\; {\rm Cov}^{-1}\br{\Pi_{\rm GI}^{(i)}(\ell),\Pi_{\rm GI}^{(j)}(\ell)}\; \frac{\partial \Pi_{\rm GI}^{(j)}(\ell)}{\partial p_\nu}\;.
} 
Note that, contrary to works focusing on cosmic shear analyses and treating intrinsic alignments as the systematic, we use the GI contribution in (\ref{eq:fisher}) and insert the transformed GG signal into (\ref{eq:bias}) such that it plays the role of a systematic. We make the assumption that, given $z_i<z_j$, the galaxy redshift distribution $p^{(i)}(z)$ entering (\ref{eq:GIdef}) is sufficiently compact that we can take the term $\bb{(1+z_i)/(1+z_{\rm piv})}^\gamma$ out of the comoving distance integration. Then both parameter derivatives needed for (\ref{eq:fisher}) and (\ref{eq:bias}) are readily calculated analytically.

\begin{figure}[t]
\centering
\includegraphics[scale=.6]{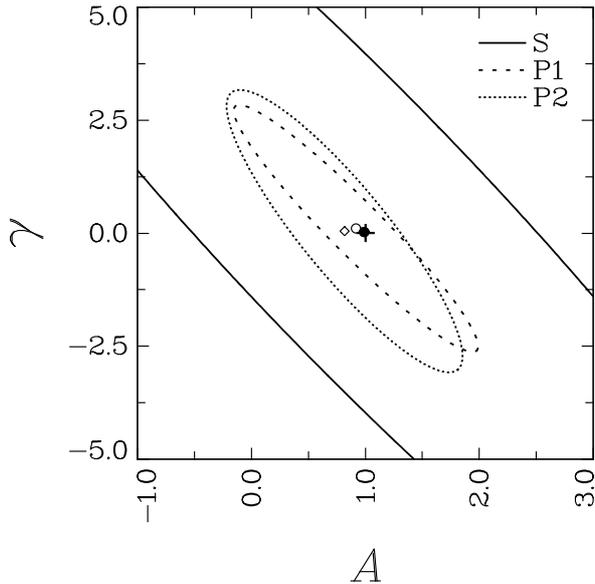}
\caption{Constraints on the free parameters of the GI model. Shown are the $1\,\sigma$ confidence contours for the different survey models used. The solid ellipse with its centre indicated by a filled circle corresponds to the spectroscopic survey S, the long-dashed ellipse and open circle to survey P1, and the short-dashed ellipse with diamond to survey P2. Note that the centres of the contours are offset due to the bias by the residual GG signal. The cross marks the fiducial values of $A=1$ and $\gamma=0$.}
\label{fig:boost_fisher}
\end{figure}


Since at this point we merely seek to demonstrate the concept of boosting, we limit the set of $\Pi^{(i)}(\ell)$ entering (\ref{eq:fisher}) to those bins $i$ which fulfil $z_i \in \bb{0.4;1.4}$. This ensures that the approximation (\ref{eq:proxyforb}) can be used throughout and that it is straightforward to assign $\sigma_{\rm opt}$. Besides, we avoid issues at high $z_i$ with non-zero $G^{(i)}(\chi(z_{\rm max}))$, which could possibly violate the basic condition $G^{(i)}(\chi_{\rm hor})=0$, see Sect.$\,$\ref{sec:solveweight}. We determine $\sigma_{\rm opt}$ by computing the diagnostic $\zeta$ given by (\ref{eq:diagnostic}) for all $z_i$ and devising simple, piecewise linear formulae which yield a $\sigma_{\rm opt}$ in the regime of small $\zeta$ for every $z_i$. For survey P2 (standard photo-z) we use $\sigma_{\rm opt} = 0.02\, z_i + 0.035$. The two other surveys have the same redshift binning and hence identical $\zeta$. We set $\sigma_{\rm opt} = 0.13\, z_i - 0.014$ for $z_i \leq 1$ and $\sigma_{\rm opt} = -0.057\, z_i + 0.173$ for $z_i > 1$ in these cases.

Now we are in the position to compute the boosting transformation for power spectra with $z_i \in \bb{0.4;1.4}$. By means of (\ref{eq:fisher}) and (\ref{eq:bias}) we obtain statistical and systematic error estimates for both intrinsic alignment parameters for all three survey models, summarised in Table \ref{tab:boost_fisher}. When varying both parameters, we find marginalised $1\,\sigma$ errors of approximately 2.9 for $A$ and 7.4 for $\gamma$ in case of survey S. The two surveys with photometric redshift data produce errors around 0.7 on $A$ and of the order 2 for $\gamma$. As expected, the bias due to the remaining cosmic shear signal is negligible in the case of the spectroscopic survey S and clearly subdominant in the case of survey P1. Even for the standard photo-z setup P2 biases remain within the statistical $1\,\sigma$ errors, reaching up to $|b_{\rm sys}/\sigma_{\rm stat}| \approx 1/4$ for $A$.

In Fig.$\,$\ref{fig:boost_fisher} the corresponding $1\,\sigma$ confidence contours in the parameter plane $A - \gamma$ are given for the three survey models. As we have chosen a pivot redshift which is below the minimum redshift of GI signals that enter the analysis, a positive $\gamma$ leads to an increase in the amplitude of the GI model, which can be compensated by a smaller $A$. Hence, $A$ and $\gamma$ are anti-correlated, leading to the degeneracy as indicated by the error ellipses. The bias acts mainly on $A$ because a residual GG signal will to zeroth order affect the overall amplitude of the signal. In all three cases the $1\,\sigma$ contours comfortably enclose the fiducial, true parameter values.

Due to the low number density of galaxies, survey S is clearly not competitive in constraints on intrinsic alignment properties. The results from the two other surveys are not capable of pinning down the intrinsic alignment model with high precision, but their bounds are comparable to current constraints by analyses of spectroscopic measurements of galaxy number density-shape cross-correlations \citep{mandelbaum09}. Note that the weights used for this analysis may still have considerable room for optimisation, and that we only used a limited range of $z_i$. 

Table \ref{tab:boost_fisher} also lists the resulting errors when only $A$ is varied and no additional redshift dependence of the intrinsic alignment model is assumed. Constraints improve significantly when lifting the degeneracy with $\gamma$ such that $A$ is determined to better than $\pm 0.3\; (1\,\sigma)$ for the survey models with photometric redshifts while constraints by survey S are about three times weaker. The bias is still negligible for the spectroscopic survey model, and clearly subdominant for survey P1 ($\sigma_{\rm ph}=0.03$). The residual systematic affects the error budget noticeably for the analysis of survey P2 ($\sigma_{\rm ph}=0.05$) with $|b_{\rm sys}/\sigma_{\rm stat}| \approx 63\,\%$. Again, optimisation of the boosting procedure may further decrease the residual cosmic shear signal well below the statistical $1\,\sigma$-limit.

The errors for the good photo-z and in particular the spectroscopic survey models are dominated by shape noise due to the low number density of galaxies in each tomographic galaxy sample, apart from only the smallest angular frequencies. As can be seen from (\ref{eq:covPtom}), the errors scale inversely with the total number of galaxies in the survey if cosmic variance is negligible. Thus, if in the future larger number densities of galaxies with highly accurate photometric redshifts than assumed in this work are attainable, the constraints on GI correlations via the boosting technique will improve accordingly. If we re-run the analysis for survey S with the galaxy number density assumed for survey P1, i.e. a factor of 10 higher, all the statistical errors indeed decrease by almost an order of magnitude.

\section{Relation to GI nulling}
\label{sec:nulling}

If one is able to extract the GI signal from cosmic shear data, the question arises whether this could also be used to remove the GI contamination from the data and thus make cosmic shear analyses robust against biases due to intrinsic alignments. Intuitively, one can simply subtract an isolated GI signal from the original measures, and indeed we are going to devise such a procedure. Afterwards we will again propose a simple, parametric weight function to construct a boosting method, whose outcome will then be used to eliminate the GI signal. These steps are not optimised and merely serve to demonstrate the link between GI boosting and its removal, as well as to compare the performance of the latter to the standard nulling technique of \citet{joachimi08b,joachimi09} in a simple scenario.

\subsection{Signal transformation}
\label{sec:variant2}

As an alternative to the procedure in Sect.$\,$\ref{sec:trafo}, one can choose the lower integration boundary in (\ref{eq:defpi}) as $\chi_{\rm min}=\chi_i$. As is evident from (\ref{eq:modGI}), in this case only the first term of the transformed GI signal remains. Hence, it is likely that $\chi_{\rm min}=0$ produces a larger amplitude of the modified GI power spectrum, but $\chi_{\rm min}=\chi_i$ results in a cleaner signal insofar as it contains only contributions from intrinsic alignments generated by matter at distance $\chi_i$. Consequently, we are going to use the latter choice of $\chi_{\rm min}$ for constructing a method to remove the GI signal at $\chi_i$. The transformed lensing signal for $\chi_{\rm min}=\chi_i$ is derived in analogy to (\ref{eq:modGG}) and reads
\eqa{
\label{eq:modGG2}
\Pi^{(i)}_{\rm GG}(\ell) &=& \frac{9H_0^4 \Omega_{\rm m}^2}{4 c^4} \int_0^{\chi_i} \dd \chi \int_{\chi_i}^{\chi_{\rm hor}} \!\!\!\! \dd \bar{\chi}\; B^{(i)}(\bar{\chi}) \br{1 - \frac{\chi}{\bar{\chi}}}\\ \nn
&& \hspace*{2cm} \times\; \br{1 - \frac{\chi}{\chi_i}} \bc{1+z(\chi)}^2 P_\delta \br{\frac{\ell}{\chi},\chi}\;.
}

Now suppose we are able to construct a boosting technique with a significant signal $\Pi_{\rm GI}^{(i)}(\ell)$ while $\Pi_{\rm GG}^{(i)}(\ell) \approx 0$. Noting again that the remaining first term in (\ref{eq:modGI}) is a rescaled version of the original GI signal (\ref{eq:GIapprox}), we define a further set of power spectra
\eq{
\label{eq:defnulling}
Q^{(ij)}_{\rm obs}(\ell) \equiv P^{(ij)}_{\rm obs}(\ell) - f_{ij}\; \Pi^{(i)}_{\rm obs}(\ell)\; ~~~~\mbox{with}~~ f_{ij} = \frac{g^{(j)}(\chi_i)}{G^{(i)}(\chi_i)}\;,
}
and likewise for the individual GG and GI signals. This definition holds for all $i < j$. The auto-correlations $Q^{(ii)}(\ell)$ would simply correspond to the original auto-correlation power spectra $P^{(ii)}(\ell)$. As we are still working in the approximation of very narrow redshift bins, auto-correlations are hardly affected by GI correlations at all. In practice, auto-correlations are likely to be excluded or specially treated anyway due to the presence of intrinsic ellipticity correlations, see the discussion in Sect.$\,$\ref{sec:conclusions}.

Assuming that the GI boosting works effectively, $\Pi_{\rm GG}^{(i)}(\ell) \approx 0$, so that one expects that $Q_{\rm GG}^{(ij)}(\ell) \approx P_{\rm GG}^{(ij)}(\ell)$, i.e. the transformed cosmic shear signal is close to the original GG term. Switching to the notation of narrow redshift bins again, we find for the transformed GI signal
\eqa{
Q_{\rm GI}(\chi_i,\chi_j,\ell) &=& \frac{3H_0^2 \Omega_{\rm m}}{2 c^2}\; g(\chi_j,\chi_i) \frac{1+z(\chi_i)}{\chi_i}\; P_{\delta {\rm I}} \br{\frac{\ell}{\chi_i},\chi_i}\\ \nn
&& \hspace*{-.5cm} - f_{ij}\; \frac{3H_0^2 \Omega_{\rm m}}{2 c^2}\; G^{(i)}(\chi_i)\; \frac{1+z(\chi_i)}{\chi_i}\; P_{\delta {\rm I}} \br{\frac{\ell}{\chi_i},\chi_i} = 0\;,
}
where we have inserted (\ref{eq:GIapprox}) and the first term of (\ref{eq:modGI}), and made use of the transition $g^{(j)}(\chi_i) \rightarrow g(\chi_j,\chi_i)$, see (\ref{eq:lensefftransition}). As a consequence, $Q_{\rm obs}^{(ij)}(\ell) \approx P_{\rm GG}^{(ij)}(\ell) - f_{ij}\, \Pi^{(i)}_{\rm GG}(\ell) \approx P_{\rm GG}^{(ij)}(\ell)$. Hence, if we can devise an effective boosting technique using $\chi_{\rm min}=\chi_i$, we immediately have a means of GI removal at our disposal via (\ref{eq:defnulling}).

 Note that the standard nulling technique as presented in \citet{joachimi08b} also makes use of the definition (\ref{eq:defpi}) with $\chi_{\rm min}=\chi_i$. The central condition in their approach is recovered in our formalism by requiring $G^{(i)}(\chi_i)=0$, which eliminates the GI signal under the same assumption of narrow redshift bins, see (\ref{eq:modGI}). For practical purposes we also switch to the discretised form of the signal transformation (\ref{eq:defpidiscrete}), using now $j_{\rm min}=i$.

\subsection{Construction of weights}
\label{sec:nullingweights}

We begin by developing again a boosting technique, now for the changed condition $\chi_{\rm min}=\chi_i$. Due to the associated change in the lower boundary of integration in (\ref{eq:defpi}), the condition to remove the GG signal is altered as well. Keeping the same approximations as used to derive (\ref{eq:conditionGint}), we now obtain from (\ref{eq:modGG2})
\eqa{
\label{eq:conditionGint2}
&& \hspace*{0cm} \int_0^{\chi_i} \dd \chi \int_{\chi_i}^{\chi_{\rm hor}}  \dd \bar{\chi}\; B^{(i)}(\bar{\chi}) \br{1 - \frac{\chi}{\bar{\chi}}} \br{1 - \frac{\chi}{\chi_i}}\\ \nn
&=& \frac{\chi_i}{2} \br{{\cal M}_1 - {\cal M}_2\; \frac{\chi_i}{3} } = 0\;,
}
where we executed the integration over $\chi$ and defined
\eq{
\label{eq:defM}
{\cal M}_\mu \equiv \int_{\chi_i}^{\chi_{\rm hor}} \dd \chi\; B^{(i)}(\chi)\; \chi^{1-\mu}\;; ~~~\mu=1,2\;.
}
Inserting (\ref{eq:volterrasolution}) into the foregoing definition and integrating by parts, one arrives at the useful relations
\eqa{
{\cal M}_1 &=& G^{(i)}(\chi_i) - \chi_i\; \frac{\partial G^{(i)}}{\partial \chi} (\chi_i)\;;\\ \nn
 {\cal M}_2 &=& - \frac{\partial G^{(i)}}{\partial \chi} (\chi_i)\;.
}
When these are plugged into (\ref{eq:conditionGint2}), we obtain a condition which is the equivalent of (\ref{eq:conditionGint}), i.e. which ensures the suppression of the GG signal in the transformed power spectra (\ref{eq:defpi}),
\eq{
\label{eq:conditionG2}
\frac{\partial G^{(i)}}{\partial \chi} (\chi_i) = \frac{3}{2\,\chi_i}\; G^{(i)}(\chi_i)\;.
}
In contrast to (\ref{eq:conditionGint}), which is an integral condition on $G^{(i)}(\chi)$ and in its discrete form involves all sampling points between $z_{\rm min}$ and $z_i$, (\ref{eq:conditionG2}) is local and even contains a derivative of $G^{(i)}(\chi)$. Hence, we suspect that (\ref{eq:conditionG2}) is less robust against the inevitable discretisation of the weight function. As a cross-check for the accuracy of (\ref{eq:conditionGint2}), and equivalently (\ref{eq:conditionG2}), we define the additional diagnostic
\eqa{
\eta &\equiv& \left| {\cal M}_1 - {\cal M}_2\; \frac{\chi_i}{3} \right|\\ \nn
&\approx& \left| \sum_{j=i}^{N_z} B^{(i)}(\chi(z_j))\; \chi'(z_j)\; \Delta z_j \br{1 - \frac{\chi(z_i)}{3\chi(z_j)}}\; \right|\;,
}
where the integrals (\ref{eq:defM}) were transformed to redshift and discretised in analogy to (\ref{eq:defpidiscrete}).

Moreover, (\ref{eq:conditionG2}) hinders us to impose the condition $\partial G^{(i)}/\partial \chi |_{\chi_i}=0$ again, which boosted the GI term, see (\ref{eq:modGI}). We define
\eq{
\label{eq:defG2}
G_Q^{(i)}(\chi) \equiv {\cal N}\; \exp \bc{-\frac{(\chi-\chi_i)^2}{\sigma^2}} \br{\chi - b}\;,
}
which has one free parameter less than (\ref{eq:defG}). To avoid any confusion with foregoing usage, we will add a sub- or superscript $Q$ to indicate quantities which are used in this section for devising a nulling procedure. The condition (\ref{eq:conditionG2}) readily implies $b=\chi_i/3$. As long as $\sigma/\chi_i \ll 1$, (\ref{eq:defG2}) has an extremum in the vicinity of $\chi_i$, located at
\eq{
\chi_{\rm extr} = \frac{2}{3}\; \chi_i + \frac{1}{3}\; \chi_i\; \sqrt{1 + \frac{9 \sigma^2}{2 \chi_i^2}}\;.
}
Therefore $G_Q^{(i)}(\chi)$ as defined in (\ref{eq:defG2}) should nonetheless boost the tranformed GI signal fairly well. In complete analogy to the derivation in Sect.$\,$\ref{sec:weights}, one obtains the weight function
\eqa{
\label{eq:weightfunction2}
B_Q^{(i)}(\chi) &=& {\cal N}\; \frac{2 \chi}{\sigma^2}\; \exp \bc{-\frac{(\chi-\chi_i)^2}{\sigma^2}}\\ \nn
&& \hspace*{1.5cm} \times\; \bc{2\, (\chi - \frac{\chi_i}{3})\; \frac{(\chi-\chi_i)^2}{\sigma^2} - 3 \chi + \frac{7}{3} \chi_i}\;.
}
The normalisation ${\cal N}$ is again given by (\ref{eq:normalisation}). As before, this weight function still has one free parameter $\sigma$ which will be used to optimise the representations of the continuous functions (\ref{eq:defG2}) and (\ref{eq:weightfunction2}) by the discrete set of sampling points entering (\ref{eq:defpi}). The weights derived from (\ref{eq:weightfunction2}) yield GI-boosted power spectra $\Pi_Q^{(k)}(\ell)$ via (\ref{eq:defpi}), which in turn produce GI-nulled measures via (\ref{eq:defnulling}).

\subsection{Nulled signals}
\label{sec:nullingperformance}

\begin{figure}[t]
\centering
\includegraphics[scale=.37,angle=270]{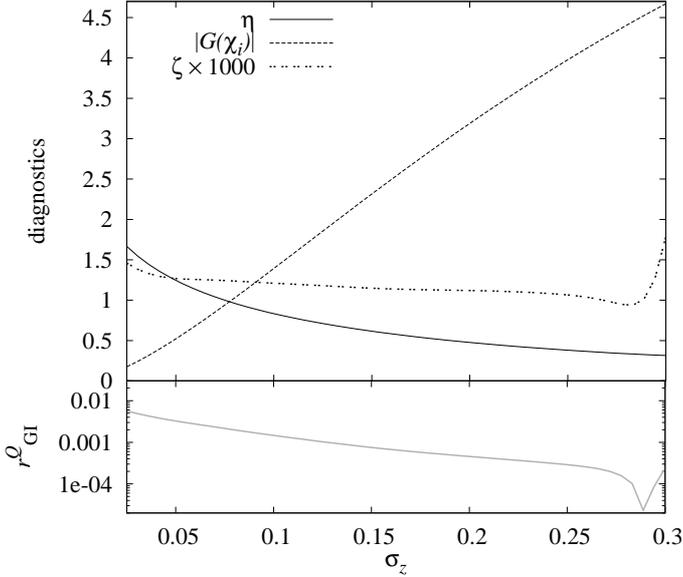}
\caption{Determination of $\sigma_{\rm opt}$ for the spectroscopic survey S at $z_i=0.53$. \textit{Top panel}: Diagnostics $\zeta$ (dotted line), $\eta$ (solid line), and $|G_Q^{(i)}(\chi_i)|$ (dashed line) as a function of $\sigma_z$. Larger $\sigma_z$ yield the desired small values of $\eta$ and larger $|G_Q^{(i)}(\chi_i)|$, in agreement with the minimum of $\zeta$ which is found at $\sigma_z \approx 0.28$. Note that we have rescaled $\zeta$ for convenience. \textit{Bottom panel}: GI over GG ratio $r^Q_{\rm GI}$ as a function of $\sigma_z$. The diagnostics indeed hint at a regime of $\sigma_z$ where $r^Q_{\rm GI}$ is smallest.}
\label{fig:sigopt2}
\end{figure}

Again we study a set of diagnostics as a function of $\sigma_z$ to identify regimes of $\sigma_z$ where the GI nulling performs well. In Fig.$\,$\ref{fig:sigopt2} we plot $\zeta$ as defined in (\ref{eq:diagnostic}), $\eta$ which assesses how (\ref{eq:conditionG2}) is affected by the discretisation, and $|G_Q^{(i)}(\chi_i)|$ as an indicator of the boosting of the GI signal in the $\Pi_Q^{(i)}(\ell)$\footnote{Note that since we have normalised $G_Q^{(i)}(\chi)$, $|G_Q^{(i)}(\chi_i)|$ is a meaningful measure of the size of $G_Q^{(i)}(\chi)$ at $\chi_i$, relative to its overall amplitude.}, for the spectroscopic survey S at $z_i=0.53$. Furthermore we show the GI over GG ratio $r^Q_{\rm GI}$, which is given by (\ref{eq:rgi}) when replacing $X$ by the nulled power spectra (\ref{eq:defnulling}). Note that small values of $r^Q_{\rm GI}$ are indicative of an effective removal of the GI signal.

One might expect that $|G_Q^{(i)}(\chi_i)|$ is largest for small $\sigma_z$ because $G_Q^{(i)}(\chi)$ is sharply peaked with a large maximum value. However, this effect is counteracted by the normalisation of the weight function. Large values of $\sigma_z$ cause $G_Q^{(i)}(\chi)$ to be smoother, i.e. to have smaller curvature. Due to (\ref{eq:volterrasolution}) the amplitude of $B_Q^{(i)}(\chi)$ would thus decrease for fixed normalisation. Since we normalise $B_Q^{(i)}(\chi)$ according to (\ref{eq:normalisation}) for every $\sigma_z$ individually, large $\sigma_z$ yield a higher normalisation relative to small $\sigma_z$, implying also larger values of $G_Q^{(i)}(\chi)$. Hence, one observes an increase in $|G_Q^{(i)}(\chi_i)|$ as a function of $\sigma_z$.

The diagnostic $\eta$ has relatively large values for strongly peaked $G_Q^{(i)}(\chi)$ and decreases slowly for larger $\sigma_z$. A small change in the weight function $B_Q^{(i)}(\chi)$ due to the discretisation can induce significant changes in $G_Q^{(i)}(\chi)$, and its slope close to $\chi_i$, which are the stronger the more sharply peaked $G_Q^{(i)}(\chi)$ is. Therefore (\ref{eq:conditionG2}) is more difficult to fulfil at small $\sigma_z$. Both $|G_Q^{(i)}(\chi_i)|$ and $\eta$ prefer larger $\sigma_z$, in agreement with $\zeta$, which we thus continue to use for the determination of $\sigma_{\rm opt}$. As $\zeta$ clearly disfavours $\sigma_z \gtrsim 0.3$, we choose as the optimum the minimum of $\zeta$ at about 0.28. Considering the lower panel of Fig.$\,$\ref{fig:sigopt2}, this value is in very good agreement with small and hence close to optimal values of $r^Q_{\rm GI}$. Generally, we find that $\sigma_{\rm opt}$ is considerably larger for this approach, compared to the variant analysed in Sect.$\,$\ref{sec:performance}.

\begin{figure}[t]
\centering
\includegraphics[scale=.44,angle=270]{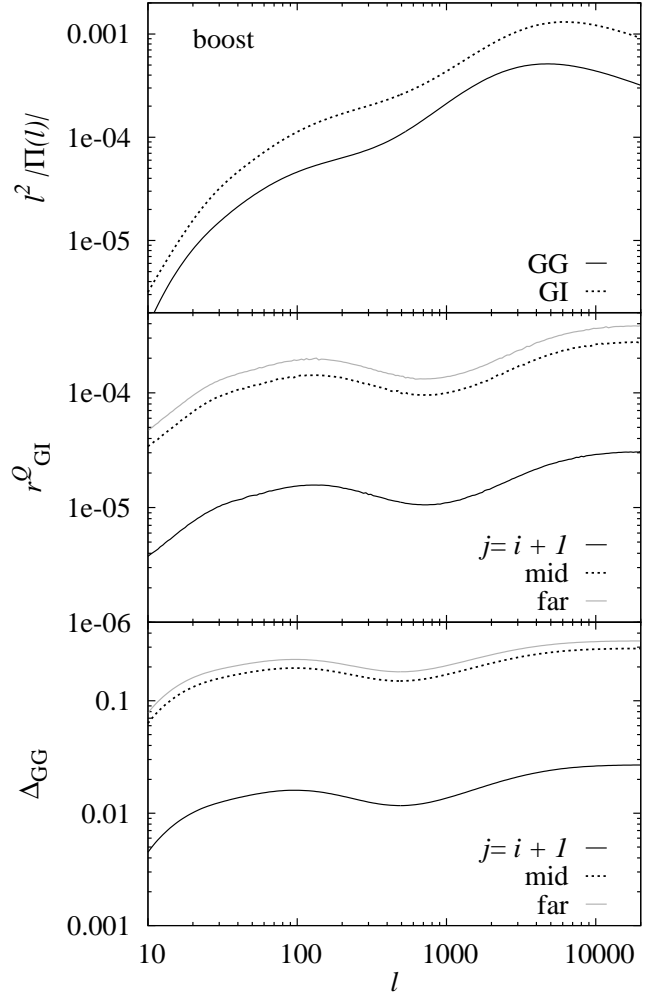}
\caption{Nulling performance for the spectroscopic survey S at $z_i=0.53$. \textit{Top panel}: Transformed GG (solid curve) and GI (dotted curve) power spectra, computed according to (\ref{eq:defpi}). \textit{Centre panel}: GI over GG ratio $r^Q_{\rm GI}$ as a function of angular frequency for the same parameters as above. The background redshift bin $j$ is set to $j=i+1$ (solid black curve), $j=(i+N_z)/2$ (\lq mid\rq; dotted black curve), and $j=N_z$ (\lq far\rq; solid grey curve). \textit{Lower panel}: Relative deviation $\Delta_{\rm GG}$ as a function of angular frequency for the same parameters as above. The coding of the curves is identical to the foregoing case.}
\label{fig:BG2}
\end{figure}

With this finding at hand, we can compute GI-boosted power spectra according to (\ref{eq:defpi}), and from these sets of nulled power spectra via (\ref{eq:defnulling}), results for both being shown in Fig.$\,$\ref{fig:BG2}. The GI term is significantly less boosted than in the version studied in Sect.$\,$\ref{sec:boostedsignals} with $r_{\rm GI}$ less than 10 (see Table \ref{tab:rgi} for comparison). Still, the intrinsic alignment suppression works excellently with $r^Q_{\rm GI} \lesssim 5 \times 10^{-4}$ for all background redshift bins and angular frequencies.

\begin{table*}[t!!!]
\centering
\caption{Summary of the nulling performance for two survey models and different values of $z_i$. Given are the median values of the GI over GG ratio of the power spectra $Q^{(ij)}(\ell)$, $r^Q_{\rm GI}$, and the relative deviation of $Q^{(ij)}(\ell)$ from the original power spectra, $\Delta_{\rm GG}$, for the correlation of adjacent bins (\lq $i\!+\!1$\rq), the cross-correlation between bin $i$ and the background bin with index $(i+N_z)/2$ (\lq mid\rq), and the cross-correlation between bin $i$ and the most distant bin with index $N_z$ at $z \lesssim 2$ (\lq far\rq). In addition $\sigma_{\rm opt}$ as determined from $\zeta$ is listed for every case considered.}
\begin{tabular}[t]{ccccccccc}
survey & $z_i$ & $\sigma_{\rm opt}$ & $r^Q_{\rm GI}(i\!+\!1)$ & $r^Q_{\rm GI}$(mid) & $r^Q_{\rm GI}$(far) & $\Delta_{\rm GG}(i\!+\!1)$ & $\Delta_{\rm GG}$(mid) & $\Delta_{\rm GG}$(far)\\
\hline\hline
S  & 0.53 & 0.280 & $0.1 \! \times \! 10^{-4}$ & $1.3 \! \times \! 10^{-4}$ & $1.7 \! \times \! 10^{-4}$ & 0.01 & 0.19 & 0.22\\
   & 0.76 & 0.255 & $0.4 \! \times \! 10^{-4}$ & $1.2 \! \times \! 10^{-4}$ & $1.6 \! \times \! 10^{-4}$ & 0.01 & 0.11 & 0.15\\
   & 0.96 & 0.235 & $0.3 \! \times \! 10^{-4}$ & $0.9 \! \times \! 10^{-4}$ & $1.4 \! \times \! 10^{-4}$ & 0.00 & 0.05 & 0.07\\
P1 & 0.53 & 0.280 & $7.9 \! \times \! 10^{-2}$ & $2.4 \! \times \! 10^{-2}$ & $3.2 \! \times \! 10^{-2}$ & 0.02 & 0.19 & 0.23\\
   & 0.76 & 0.255 & $4.2 \! \times \! 10^{-2}$ & $2.3 \! \times \! 10^{-2}$ & $3.0 \! \times \! 10^{-2}$ & 0.01 & 0.12 & 0.15\\
   & 0.96 & 0.235 & $2.9 \! \times \! 10^{-2}$ & $2.1 \! \times \! 10^{-2}$ & $2.8 \! \times \! 10^{-2}$ & 0.00 & 0.05 & 0.08\\
\end{tabular}
\label{tab:rginull}
\end{table*}

In Table \ref{tab:rginull} values of $r^Q_{\rm GI}$ for other $z_i$ and in addition for the good photo-z survey P1 are listed. The downweighting of the GI signal quickly deteriorates with the increase in photometric redshift uncertainty, being more than two orders of magnitude larger for survey P1. For the standard photo-z case we find that the boosting as implemented in this section is ineffective, so that we do not consider it here. As shown in Sect.$\,$\ref{sec:variant2}, under idealistic circumstances one expects the signal in the nulled power spectra $Q^{(ij)}(\ell)$ to be close to the one in the original power spectra $P^{(ij)}(\ell)$. Hence, we calculate the quantity
\eq{
\Delta_{\rm GG} \equiv \frac{Q_{\rm GG}^{(ij)}(\ell)}{P_{\rm GG}^{(ij)}(\ell)} - 1\;,
}
which is also given in Fig.$\,$\ref{fig:BG2} and Table \ref{tab:rginull}. The deviation from the original signal is at the per cent level for close foreground and background redshift bins with $z_i \lesssim z_j$, and increases to about $20\,\%$ if bins $i$ and $j$ are far apart, irrespective of the photometric redshift quality.

\subsection{Information content}
\label{sec:nullinginformation}

How does the nulling technique as outlined above perform in comparison with the standard nulling approach? For a very dense binning in redshift both methods evidently remove the GI contamination of the cosmic shear signal to high accuracy, see for instance the recent findings by \citet{shi10}. However, \citet{joachimi09} have shown that, even in idealistic situations comparable to our spectroscopic survey, a substantial loss of cosmological information is inherent to standard nulling. We assess the information content in both nulling approaches in a simple case study.

We restrict ourselves to the spectroscopic survey model S and consider again only $z_i \in \bb{0.4;1.4}$, for the same reasons as discussed in Sect.$\,$\ref{sec:constraints}. Again, we compute $\zeta$ for all $z_i$ to find a simple prescription for $\sigma_{\rm opt}$; in this case we use $\sigma_{\rm opt} = -0.131\, z_i + 0.346$. The information content is quantified in terms of the cumulative signal-to-noise (S/N), defined as 
\eq{
\label{eq:sn}
\frac{\rm S}{\rm N} = \sum_\ell \sum_{j > i,\, l > k} Q_{\rm GG}^{(ij)}(\ell)\; {\rm Cov}^{-1}\br{Q_{\rm GG}^{(ij)}(\ell),Q_{\rm GG}^{(kl)}(\ell)}\; Q_{\rm GG}^{(kl)}(\ell)\;,
}
where the covariance of the nulled power spectra can be derived from (\ref{eq:defnulling}),
\eqa{
{\rm Cov}\br{Q^{(ij)}(\ell),Q^{(kl)}(\ell)} &=& {\rm Cov}\br{P^{(ij)}(\ell),\; P^{(kl)}(\ell)}\\ \nn
&& \hspace*{-2.5cm} -\; f_{ij}\; {\rm Cov}\br{\Pi_Q^{(i)}(\ell),\; P^{(kl)}(\ell)} - f_{kl}\; {\rm Cov}\br{P^{(ij)}(\ell),\; \Pi_Q^{(k)}(\ell)}\\ \nn
&& \hspace*{-2.5cm} +\; f_{ij}\; f_{kl}\; {\rm Cov}\br{\Pi_Q^{(i)}(\ell),\; \Pi_Q^{(k)}(\ell)}\;.
}
The S/N for data sets of original power spectra $P^{(ij)}(\ell)$ and of nulled power spectra obtained via the \citet{joachimi09} formalism are calculated in analogy to (\ref{eq:sn}). For this setup it is safe to assume that (\ref{eq:covPtom}) has only contributions from shape noise. Even with this simplification, the inversion of the covariance is computationally expensive for the total of 65 bins between $z_i=0.4$ and $z_i=1.4$. Thus we include by default only tomographic measures for every fifth bin $i$, but all $j>i$, in the S/N. The absolute value of the S/N depends of course on how many power spectra are incorporated, but we are only interested in the ratio of S/N for the nulled datasets over the set of original power spectra.

Note that for every $z_i$ one can make use of $N_z-i$ power spectra $Q^{(ij)}(\ell)$. The very same number of modes is available in the standard nulling approach although one mode is discarded to perform the actual nulling \citep[for details see][]{joachimi09}. Transformed auto-correlation power spectra with $i=j$ do not enter the S/N, but by construction the $P^{(ii)}(\ell)$ do contribute to all $Q^{(ij)}(\ell)$ via the $\Pi_Q^{(i)}(\ell)$, whereas in standard nulling auto-correlations are completely discarded. However, due to the dense redshift binning, we expect the amount of independent information contained in auto-correlation power spectra to be small.

We have given the resulting ratios of the S/N for the nulled data set over the S/N for the original one in Table \ref{tab:snrnull}. The considerable loss of information can be confirmed, the S/N for both nulling methods yielding less than $20\,\%$ of the original S/N. We find that these numbers are very robust against changes in the number and values of redshift bins $i$ included in the S/N by varying the size of steps in bin numbers $i$ and the range of redshifts considered. It is quite remarkable that the ratios for both nulling methods are very similar. The slightly bigger number for the nulling as devised in this work could be related to the inclusion of auto-correlation power spectra, but is not very significant anyway. 

In the standard nulling case the information loss is caused by discarding part of the signal, namely one mode per bin $i$ whereas the variant suggested here features a signal that deviates by at most about $20\,\%$ from the untransformed one. In the latter case the loss is caused by an increase in the covariance due to the subtraction of signals in (\ref{eq:defnulling}). We conjecture at this point that the agreement in the amount of information lost, in spite of the largely different mechanisms of the two methods, hints at a fundamental limit of how far GI and GG signals can be distinguished by only relying on the redshift dependence of the two contributions.

\begin{table}[t]
\begin{minipage}[t]{\columnwidth}
\centering
\caption{Ratio of cumulative signal-to-noise of the nulled set of power spectra over the original set of power spectra (SNR). The results for the nulling method devised in this work and the standard nulling technique \citep{joachimi09} are compared. The default SNR shown is computed for a step size in foreground redshift bin index $i$ of 5 and in the redshift range $\bb{0.4;1.4}$. In addition the moduli of the fractional deviation of the SNR from the default when using a step size of 3, i.e. every third bin $i$ (third column), a step size of 7 (fourth column), and the default step size but a different redshift range $\bb{0.5;1.3}$ (fifth column) are given.}
\begin{tabular}[t]{ccccc}
nulling type & SNR & step 3 & step 7 & $z \in \bb{0.5;1.3}$\\
\hline\hline
this work & 0.179 & $0.76\,\%$ & $2.75\,\%$ & $1.96\,\%$\\
standard  & 0.163 & $0.59\,\%$ & $2.32\,\%$ & $1.45\,\%$\\
\end{tabular}
\label{tab:snrnull}
\end{minipage}
\end{table}

\section{Conclusions}
\label{sec:conclusions}

In this paper we presented a method which extracts shear-ellipticity correlations (the GI signal) from a tomographic cosmic-shear data set. The approach relies neither on models of intrinsic alignments nor on knowledge of the cosmological parameters that characterise the cosmic shear (GG) signal, making only use of the typical and well-understood redshift dependencies of both the GI and GG term. We derived constraints which a linear transformation of second-order cosmic shear measures has to fulfil in order to boost the GI signal and simultaneously suppress the lensing contribution. We studied in depth a particular parametrisation of the weights entering this transformation and analysed the performance of the resulting GI boosting technique for three representative survey models.

Applying the GI boosting to future all-sky cosmic shear surveys, it should be possible to isolate the GI signal with subdominant biases due to a residual GG term, and with constraints that are comparable to current results from indirect measurements of shear-ellipticity correlations \citep{mandelbaum09}. If one restricts the analysis to galaxies with photometric redshift information of good quality, i.e. a redshift scatter of not more than $\sigma_{\rm ph}(1+z)$ with $\sigma_{\rm ph}=0.03$, one can achieve $1\,\sigma$-errors on the GI signal amplitude $A$ in the parametrisation of (\ref{eq:GImodel}) of better than 0.2 when varying only the amplitude, and a marginalised error of approximately 0.7 when fitting an additional redshift dependence. 

Using all galaxies from a survey fulfilling $\sigma_{\rm ph} \leq 0.05$, the statistical constraints degrade only marginally but the parameter bias due to the residual GG contribution can attain more significant values of up to $b_{\rm sys}/\sigma_{\rm stat} \lesssim 2/3$. We also considered a survey with high-quality photometric or spectroscopic redshifts. However, the expected low number density of galaxies of $n_{\rm g}=1\,{\rm arcmin}^{-2}$, even for future surveys, does not permit us to place competitive constraints on intrinsic alignment models. In this case of highly accurate redshift information the residual bias on parameters is negligible.

Although we have modelled scatter in photometric redshifts for our investigations, we did not consider other effects affecting the accuracy of redshift information, such as an error in the median of the galaxy redshift distributions or catastrophic failures in the determination of photometric redshifts. As several studies of intrinsic alignment removal techniques have demonstrated \citep[e.g.][]{bridle07,joachimi09,joachimi10}, the ability to separate the GI from the GG signal depends vitally on these parameters characterising the accuracy of and knowledge about redshifts. The same can be expected for the GI boosting technique, possibly to an even larger extent since in this case one attempts to suppress the originally strongest contribution to ellipticity correlations, the GG signal. Hence, we hypothesise that the requirements of future ambitious weak lensing surveys, like a negligible fraction of catastrophic failures and an error in the mean of each redshift distribution of not more than $0.002(1+z)$ \citep{laureijs09}, are both necessary and sufficient for a success of GI boosting. We leave a detailed assessment of the requirements on the quality of redshift information to future work.

Moreover, we did not yet include intrinsic ellipticity correlations (II) into our considerations. Since the II signal is generated by physically close pairs of galaxies, it has a redshift dependence that is clearly distinct from the GI and GG terms, and can thus be removed relatively easily \citep{king02,king03,heymans03,takada04b}. In tomographic cosmic shear data it mainly affects auto-correlations and cross-correlations of adjacent photometric redshift bins with significant overlap of their corresponding distributions of true redshifts. One of the aforementioned II removal techniques could precede the GI boosting, causing an increased shape noise contribution in particular in the auto-correlations due to the reduced number of available galaxy pairs. Alternatively, the downweighting of the II signal could also be readily incorporated into the boosting technique by introducing the additional condition $\partial^2 G^{(i)}/ \partial \chi^2 |_{\chi_i} = 0$, implying $B^{(i)}(\chi_i)=0$ and therefore a downweighting of auto-correlations as well as cross-correlations of adjacent redshift distributions, see (\ref{eq:defpi}).

Our findings still have the potential for significant improvement because we have only considered one specific parametrisation of the weight function that governs the boosting transformation. While this choice is intuitive and allows analytical progress, a more versatile approach could be to assume the weight function $B^{(i)}(\chi)$ as piecewise linear, with nodes placed at the median redshift of every galaxy redshift sample. The constraints on GI boosting and GG suppression could then be directly imposed on the discretised version of the boosting transformation, thereby fixing a subset of the values of $B_Q^{(i)}(\chi)$ at its nodes. The remaining freedom in the weight function could for instance be used to maximise the signal-to-noise of the expected transformed GI signal.

We also constructed a method of GI removal, directly based on a slightly modified version of the GI boosting technique. In principle, we showed that if one is able to isolate the GI signal via boosting, one can simply subtract a rescaled version of the GI term from the original cosmic shear measures to eliminate the intrinsic alignment systematic. We find that the residual contamination of the cosmic shear signal by GI correlations is indeed small, and that the cumulative signal-to-noise of the thus treated cosmic shear signal decreases by about a factor of 6. This value is remarkably close to the result for the standard GI nulling technique as introduced by \citet{joachimi08b,joachimi09}, in spite of the differing approaches. The underlying reason for this agreement may be due to a fundamental limit in the ability to separate GI and GG signals relying only on the dependence on redshift, which is worth to be addressed in future investigations. Of course, such a limit would also imply a maximum accuracy with which parameters of intrinsic alignments can be constrained via GI boosting.

Like the method devised in this work, the standard nulling technique is also a purely geometrical method. Hence, a combined application of GI boosting and nulling to a cosmic shear data set would still be based on a minimum of assumptions about the actual forms of signals or the values of model parameters. For instance one could use an initial analysis based on nulling to yield robust estimates of the cosmic shear signal and the corresponding cosmological model. This could then be used to construct weights for the GI boosting transformation such that even in the case of standard photometric redshift quality (which we assumed to be $\sigma_{\rm ph}=0.05$ in this paper) the bias due to the residual GG signal would be negligible, thereby enabling an equally robust estimate of the GI signal.

Ultimately, the cosmic shear analysis, the treatment of intrinsic alignments, and the inclusion of additional information from galaxy number density correlations \citep[as in][]{mandelbaum06,hirata07,mandelbaum09} will all be efficiently combined into a simultaneous analysis of the form presented in \citet{bernstein08} and \citet{joachimi10}, provided one can summon the computational power. Yet the model-independent, direct, and robust boosting technique, as well as nulling and the combination of the two, will prove useful e.g. to provide reliable priors on the large set of parameters entering the integrative approaches and in addition serve as a valuable consistency check in cosmic shear analyses.

\begin{acknowledgements}
We would like to thank Filipe Abdalla, Adam Amara, Sarah Bridle, and Tom Kitching for many helpful discussions on intrinsic alignments. Moreover we are grateful to our referee for a helpful report. BJ acknowledges support by the Deutsche Telekom Stiftung and the Bonn-Cologne Graduate School of Physics and Astronomy. This work was supported by the RTN-Network \lq DUEL\rq\ of the European Commission, and the Deutsche Forschungsgemeinschaft under the Transregional Research Center TR33 \lq The Dark Universe\rq.
\end{acknowledgements}

\bibliographystyle{aa}

\end{document}